\documentclass[11pt,english]{article}
\usepackage[T1]{fontenc}
\usepackage[latin9]{inputenc}
\usepackage{verbatim}
\usepackage{float}
\usepackage{amsmath}
\usepackage{graphicx}
\usepackage{esint}
\usepackage{hyperref}

\makeatletter

\providecommand{\tabularnewline}{\\}

\usepackage{latexsym}
\usepackage{epsfig}
\usepackage{enumerate}
\renewcommand{\theequation}{\thesection.\arabic{equation}}

\allowdisplaybreaks
\usepackage[numbers,sort&compress]{natbib}
\numberwithin{equation}{section}
\makeatother

\usepackage{babel}
\begin{document}
{}~ \hfill\vbox{\hbox{CTP-SCU/2013003}}\break \vskip 3.0cm
\centerline{\Large \bf Double Field Theory Inspired Cosmology}
\vspace*{10.0ex}
\begin{center}
 Houwen Wu and Haitang Yang
\end{center}
\vspace*{7.0ex}

\vspace*{4.0ex}

\centerline{\large \it Center for theoretical physics}
\centerline{\large \it Sichuan University} \centerline{\large \it
Chengdu, 610064, China}

\vspace*{1.0ex}

\centerline{E-mail: 2013222020003@stu.scu.edu.cn, hyanga@scu.edu.cn}

\vspace*{10.0ex}

\centerline{\bf Abstract}
\bigskip
Double field theory proposes a generalized spacetime action
possessing manifest T-duality on the level of component fields. We
calculate the cosmological solutions of double field theory with
vanishing Kalb-Ramond field. It turns out that double field theory
provides a more consistent way to construct cosmological solutions
than the standard string cosmology. We construct solutions for vanishing and non-vanishing symmetry preserving dilaton potentials. The solutions assemble
the pre- and post-big bang evolutions in one single line element.
This novel feature provides a natural way for the theory to contain extra dimensions. Our results show a smooth evolution from an anisotropic early stage to an isotropic phase without any special initial conditions in contrast to previous models. In addition, we demonstrate that the contraction of the dual space automatically leads to both an inflation phase and a decelerated expansion of the ordinary space during different evolution stages.

\smallskip

\vfill \eject \baselineskip=16pt \vspace*{10.0ex}

\tableofcontents


\section{Introduction}
Double field theory (DFT) was first proposed to realize T-duality
explicitly at the level of component fields of closed string field
theory \cite{Hull:2009mi,Hohm:2010pp,Siegel:1993xq}. Earlier efforts are traced back to
\cite{Tseytlin:1990nb, Duff:1989tf}.  The
equivalence of spacetime momenta and winding numbers in the string
spectra leads to an introduction of a set of dual coordinates
$\tilde x^i$, conjugated to winding numbers. These dual
coordinates are treated on the same footing as the usual
coordinates $x^i$. The dimensionality of spacetime is then doubled
from $D$ to $D+D$. T-duality is manifested as an
$O\left(D,D\right)$ symmetry in the action. The full set of
coordinates are denoted as
$X^{M}=\left(\tilde{x}_{i},x^{i}\right)$, where $M=1,2,\ldots,2D$
is the $O\left(D,D\right)$ index, $x^{i}$ ($i=1,2,\ldots,D$) is
the usual spacetime coordinate and  $\tilde{x}_{i}$ represents the
dual coordinate. All the spacetime component fields are dependent
on both the usual and the dual coordinates: $\phi_I
\left(\tilde{x}_{i},x^{i}\right)$.

A full construction of DFT is still an open question. The current
research focuses on the massless sector of closed string spectra.
The field content includes a $D$ dimensional metric $g_{ij}$, a
scalar dilaton $\phi$ and the anti-symmetric Kalb-Ramond field
$b_{ij}$. The manifestly $O\left(D,D\right)$ invariant action
built on the generalized $O\left(D,D\right)$ metric tensor
$\mathcal{H}^{MN}$ is

\begin{eqnarray}
S & = & \int dxd\tilde{x}e^{-2d}\left(\frac{1}{8}\mathcal{H}^{MN}\partial_{M}
\mathcal{H}^{KL}\partial_{N}\mathcal{H}_{KL}-\frac{1}{2}\mathcal{H}^{MN}
\partial_{N}\mathcal{H}^{KL}\partial_{L}\mathcal{H}_{MK}\right.\nonumber \\
 &  & \left.-\partial_{M}d\partial_{N}\mathcal{H}^{MN}+4\mathcal{H}^{MN}
 \partial_{M}d\partial_{N}d\right),
\label{DFTS}
\end{eqnarray}
where the dilaton $d= \phi - \frac{1}{4}\ln g$ is an $O\left(D,D\right)$ scalar. The
indices $M,N,\cdots$ are raised and lowered by the off-diagonal
$O\left(D,D\right)$ metric and are all contracted. The component
fields $g^{ij}$ and $b^{ij}$ enter the action via the definition
\begin{equation}
\mathcal{H}_{MN}=\left(%
\begin{array}{cc}
  g^{ij} & -g^{ik}b_{kj} \\
  b_{ik}g^{kj} & g_{ij}-b_{ik}g^{kl}b_{lj} \\
\end{array}%
\right). \label{GMH}
\end{equation}

\noindent The level matching condition in
closed string theory imposes the \emph{weak constraint}
$\partial\tilde\partial \phi(x,\tilde x)=0$ for an arbitrary field
$\phi(x,\tilde x)$. The action is invariant under the gauge
transformation
\begin{eqnarray}
\delta_{\xi}\mathcal{H}^{MN} & = & \hat{\mathcal{L}}_{\xi}\mathcal{H}^{MN}\equiv\xi^{P}
\partial_{P}\mathcal{H}^{MN}+\left(\partial^{M}\xi_{P}-\partial_{P}\xi^{M}\right)
\mathcal{H}^{PN}+\left(\partial^{N}\xi_{P}-\partial_{P}\xi^{N}\right)\mathcal{H}^{MP}\nonumber \\
\delta d & = &
\xi^{M}\partial_{M}d-\frac{1}{2}\partial_{M}\xi^{M},
\end{eqnarray}
with gauge parameters
$\xi^{M}=\left(\tilde{\xi}_{i},\xi^{i}\right)$ and ``generalized
Lie derivatives'' $\hat{\mathcal{L}}_{\xi}$. The gauge
transformation $\xi^{i}$ is the traditional diffeomorphism and
$\tilde\xi^i$ is the dual diffeomorphism. To ensure the action is locally equivalent to the low energy effective string action,  a \emph{strong constraint} is needed:
$\partial\tilde\partial =0$ as an operator equation, acting on any
products of the fields. This strong constraint is also
sufficient in the construction of DFT based on closed string
field theory beyond cubic order. The  low energy effective action
of closed string theory is

\begin{equation}
S_{*}=\int
d^{D}x\sqrt{-g}e^{-2\phi}\left[R+4\left(\partial_{\mu}\phi\right)^{2}-
\frac{1}{12}H_{ijk}H^{ijk}\right],
\end{equation}
where $R$ is the Ricci scalar constructed from the string metric
$g_{\mu\nu}$, $\phi$ is the usual diffeomorphic dilaton and
$H_{ijk}=3\partial_{\left[i\right.}b_{\left.jk\right]}$ is the
field strength of the Kalb-Ramond $b_{ij}$ field. This action,
also named as tree-level string action, is the foundation of tree
level string cosmology.

Since the pioneer work of Hull and Zwiebach \cite{Hull:2009mi},
many progresses have been achieved. Good reviews are referred to
\cite{Zwiebach:2011rg,Aldazabal:2013sca} and various developments
can be found in \cite{Kwak:2010ew}-\cite{Hohm:2013jaa}. However,
to our knowledge, solutions of the action (\ref{DFTS}) have not
been constructed. The main purpose of this paper is to find
cosmological solutions of DFT. In the traditional string
cosmology, various solutions are constructed by the scale factor
duality and time reversal symmetry. However, all the solutions are
self-contained. There exists no natural way to combine two
solutions together to cover the whole spacetime of the pre- and
post-big bangs.  It is of interest to note
that the scale factor duality is an intrinsic property of DFT when
applied to a FRW like metric. This observation makes it possible
to include the pre- and post-big bangs in one single line element.
We will show that this unification of the pre- and post-big bangs manifests the existence of extra dimensions. Moreover, we demonstrate that the universe starts from an visibly anisotropic phase in the pre-big bang, evolves to an isotropic big bang and continues to an asymptotically isotropic post-big bang. This whole process needs no free parameters.

In order to simplify the story, in this paper, we set the
Kalb-Ramond field $b_{ij}=0$.\footnote{In a follow-up work, we
will incorporate non-vanishing $b$ field. From the off-diagonal
non-vanishing components in the generalized metric (\ref{GMH}),
one can expect the appearance of cross terms $dx^id\tilde x^j$.
These cross terms can not be found by the traditional string
cosmology and are novel.} We suppose the line element is FRW like

\begin{eqnarray}
dS^{2} & = & \tilde{g}^{ij}d\tilde{x}_{i}d\tilde{x}_{j}+g_{ij}dx^{i}dx^{j}\nonumber \\
 & = & -d\tilde{t}^{2}+\tilde{a}\left(t,\tilde{t}\right)^{2}\delta^{ij}d\tilde{x}_{i}d
 \tilde{x}_{j}\nonumber \\
 &  & -dt^{2}+a\left(t,\tilde t\right)^{2}\delta_{ij}dx^{i}dx^{j},
\label{MetricAnsatz}
\end{eqnarray}
where we put bars on $\tilde g^{ij}=g^{ij}$ to remind us that it
is related to $\tilde x$ for convenience of our calculation. This
notation is introduced in section 3.2.  One can see that the
non-vanishing components of the generalized metric
$\mathcal{H}_{MN}$ in (\ref{GMH}) are only $g_{ij}$ and $g^{ij}$.
Therefore, we conclude $\tilde a(t,\tilde t) = a^{-1}(t,\tilde
t)$.

There are three dilatons in double field theory. The $O(D,D)$
scalar dilaton $d$ is invariant under $O(D,D)$ transformation. The
traditional diffeomorphic scalar dilaton $\phi$ is invariant under
gauge transformation $\xi$. The dual diffeomorphic scalar dilaton
$\tilde\phi$ is invariant under dual gauge transformation
$\tilde\xi$. With the metric (\ref{MetricAnsatz}), the
relationship between the three dilatons is
\begin{eqnarray}
&&d=\phi-\frac{D-1}{2} \ln a,\nonumber\\
&&\phi=\tilde\phi +(D-1)\ln a. \label{ThreeD}
\end{eqnarray}
We will show that the dilaton $d$ is precisely the ``shifted
dilaton'' in string cosmology. The second equation of
(\ref{ThreeD}) is nothing but the scale factor duality of the
dilatons in the standard string cosmology.

Before devoted to calculations, we clarify that the continuous $O(D,D)$ symmetry is a very  fundamental symmetry. When we compactify $d$ dimensions of $D=n+d$, this symmetry breaks to $O\left(n,n\right)\times O\left(d,d;Z\right)$, where $O\left(n,n\right)$ is still a continuous group and $O\left(d,d;Z\right)$  represents T-duality in the compactified background. Since no compactification presents in string cosmology, the scale factor duality in string cosmology is not T-duality
but a realization of the continuous $O(D,D)$ symmetry specifically for the FRW metric. The $O(D,D)$ symmetry enables us to easily find solutions of DFT from these of string cosmology \cite{Sen:1991cn}.

The main purpose of this paper is to put forward cosmological solutions of DFT and discuss their physical interpretations for two scenarios. We first start from the action
(\ref{DFTS}). After substituting the metric ansatz (\ref{MetricAnsatz}) into the equations of motion (EOM) derived from the DFT action (\ref{DFTS}), we obtain two distinct metrics, the pre-big bang metric
$dS_{pre}^{2}$ for $t<0$ and the post-big bang metric $dS_{post}^{2}$ for $t>0$.  Each of the solutions consists of a pair of $O(D,D)$ connected solutions of string cosmology. This is different from the story in string cosmology, where the four line elements are completely disconnected. Both solutions unify contracting and expanding dimensions. To cover the whole spacetime and have a clearer physical picture, we introduce an $O(D,D)$ invariant dilaton potential to smooth out the singularity. We then have a unique line element describing the whole geometry, from the far past pre-big bang to the post-big bang, in contrast to string cosmology where two disjointed metrics exist. Remarkably, in both scenarios, $V(d)=0$ and $V(d)\not= 0$, the solutions manifest the existence of extra dimensions. Moreover, we explicitly show that the solutions reveal an intrinsic evolution of the universe from an isotropic pre-big bang phase to an anisotropic big bang and then to an isotropic post-big bang phase. All these new features originate from the $O(D,D)$ symmetry of the theory. This is consistent with the modern point of view that symmetries dictate physics. We further demonstrate that both the inflation and decelerated expansion of space can be triggered by the contraction of the dual space.

The reminder of this paper is outlined as follows. In section $2$,
we give a brief review on the relevant results we need in string
cosmology. Section $3$ refers to the EOM of the generalized double
action for generic metrics and the FRW like metric. In section
$4$, we give the cosmological solutions of double field theory with vanishing and non-bashing dilaton potential.
Section $5$ is our conclusion and discussions. We put some details
of the calculation in the Appendix.

\section{A brief review of string cosmology}

Since our discussions are closely related to the standard string
cosmology, in this section, we briefly review the tree level
results in string cosmology. A comprehensive treatment is referred
to \cite{Gasperini:2007zz}, on which our review is based and
references therein.

We start with the tree level string action. For the reason of
simplicity, we only consider the gravi-dilaton system without any
matter sources. The anti-symmetric Kalb-Ramond field $b_{ij}$ is
set to vanish. The action is given by

\begin{equation}
S=\frac{1}{2\kappa^2}\int
d^{D}x\sqrt{-g}e^{-2\phi}\left[R+4\left(\partial_{\mu}\phi\right)^{2}\right],
\label{SigmaS}
\end{equation}
\noindent where $D$ is the spacetime dimension, $\phi$ represents
the dilaton which is a function of $t$, and $g_{\mu\nu}$ is the
string metric. Note that the string metric is related to Einstein
metric by
$g_{\mu\nu}^{E}=\exp\left(-\frac{4}{d-1}\phi\right)g_{\mu\nu}$.
The EOM are

\begin{eqnarray}
R_{\mu\nu}+2\nabla_{\mu}\nabla_{\nu}\phi & = & 0,\nonumber \\
\nabla^{2}\phi-2\left(\partial_{\mu}\phi\right)^{2} & = & 0.
\label{EOMR}
\end{eqnarray}

\noindent An isotropic metric is adopted to study string
cosmology

\begin{equation}
ds^{2}=-dt^{2}+a\left(t\right)^{2}\delta_{ij}dx^{i}dx^{j},
\hspace{5mm} H(t)\equiv \frac{\dot a(t)}{a(t)}.
\end{equation}

\noindent With this metric, from (\ref{EOMR}), the EOM of the
graviton and the dilaton take the form

\begin{eqnarray}
\ddot{\phi}-2\dot{\phi}^{2}+\left(D-1\right)H\dot{\phi} & = & 0,\nonumber \\
2\ddot{\phi}-\left(D-1\right)\left(\dot{H}+H^{2}\right) & = & 0,\nonumber \\
\dot{H}+\left(D-1\right)H^{2}-2H\dot{\phi} & = & 0. \label{EOMCom}
\end{eqnarray}

\noindent In the convention of string cosmology, it is convenient
to introduce the ``shifted dilaton''

\begin{equation}
\psi=2\phi-\left(D-1\right)\ln a \label{ShiftedD}
\end{equation}

\noindent We will show that  this ``shifted dilaton'' $\psi$ is
the $O\left(D,D\right)$ scalar dilaton $2d$ in double field
theory. Therefore, (\ref{EOMCom}) becomes

\begin{eqnarray}
2\ddot{\psi}-\dot{\psi}^{2}-\left(D-1\right)H^{2} & = & 0,\nonumber \\
-\left(D-1\right)H^{2}+\ddot{\psi} & = & 0,\nonumber \\
\dot{H}-\dot{\psi}H & = & 0. \label{EOMComV}
\end{eqnarray}

\noindent These equations are invariant under the transformation
of the famous \emph{scale factor duality}

\begin{equation}
a\rightarrow\tilde a= a^{-1},\qquad\phi\rightarrow \tilde\phi
=\phi-\left(D-1\right)\ln a, \label{SFDuality}
\end{equation}
which leads to
\begin{equation}
\tilde{H}=-H,\qquad\tilde{\psi}=2\tilde{\phi}-\left(D-1\right)\ln
a^{-1}=\psi.
\end{equation}

\noindent There are only two independent equations in
(\ref{EOMComV}). After applying the scale factor duality on the
solution of (\ref{EOMComV}), one hence obtains

\begin{equation}
1:\qquad
a\left(t\right)=\left(\frac{t}{t_{0}}\right)^{1/\sqrt{D-1}},\qquad\psi
=-\ln\left(\frac{t}{t_{0}}\right),
\end{equation}

and its dual solution

\begin{equation}
2:\qquad\tilde{a}\left(t\right)=\left(\frac{t}{t_{0}}\right)^{-1/
\sqrt{D-1}},\qquad\tilde{\psi}=-\ln\left(\frac{t}{t_{0}}\right).
\end{equation}

\noindent The equations (\ref{EOMComV}) also possesses a ``time
reversal'' symmetry $t\rightarrow-t$. Therefore, there are two
more solutions

\begin{equation}
3:\qquad a\left(-t\right)=\left(-\frac{t}{t_{0}}\right)^{1/\sqrt{D-1}},\qquad\psi=-\ln\left(-\frac{t}{t_{0}}\right)
\end{equation}

and

\begin{equation}
4:\qquad\tilde{a}\left(-t\right)=\left(-\frac{t}{t_{0}}\right)^{-1/\sqrt{D-1}},\qquad\tilde{\psi}=-\ln\left(-\frac{t}{t_{0}}\right)
\end{equation}

\noindent In summary, four branches are found. The properties of
the solutions are listed in Table $1$.

\begin{center}
\begin{tabular}{|c|c|c|c|c|}
\hline
1 & $\dot{a}\left(t\right)>0$, expansion & $\ddot{a}\left(t\right)<0$, decelerated & $\dot{H}<0$, decreasing curvature & post-big bang\tabularnewline
\hline
2 & $\dot{\tilde{a}}\left(t\right)<0$, contraction & $\ddot{\tilde{a}}\left(t\right)>0$, decelerated & $\dot{\tilde{H}}>0$, decreasing curvature & post-big bang\tabularnewline
\hline
3 & $\dot{a}\left(-t\right)<0$, contraction & $\ddot{a}\left(-t\right)<0$, accelerated & $\dot{H}<0$, increasing curvature & pre-big bang\tabularnewline
\hline
4 & $\dot{\tilde{a}}\left(-t\right)>0$, expansion & $\ddot{\tilde{a}}\left(-t\right)>0$, accelerated & $\dot{\tilde{H}}>0$, increasing curvature & pre-big bang\tabularnewline
\hline
\end{tabular}
Table $1$. The properties of the solutions in tree level string
cosmology.
\par\end{center}

Note that deceleration occurs when
$\mathrm{sign\;}\dot{a}=-\mathrm{sign}\;\ddot{a}$, acceleration
occurs when $\mathrm{sign\;}\dot{a}=\mathrm{sign}\;\ddot{a}$. When
$H^{2}$ or $\tilde{H}^{2}$ is growing with time, the curvature is
increasing, otherwise, the curvature is decreasing. Moreover, when
$H>0$, the universe is expanding, otherwise, the universe is
contracting. All these solutions share the curvature singularity
located at $\left|t\right|\rightarrow0$, as illustrated in Figure
$1$.

\begin{figure}[H]
\begin{centering}
\includegraphics{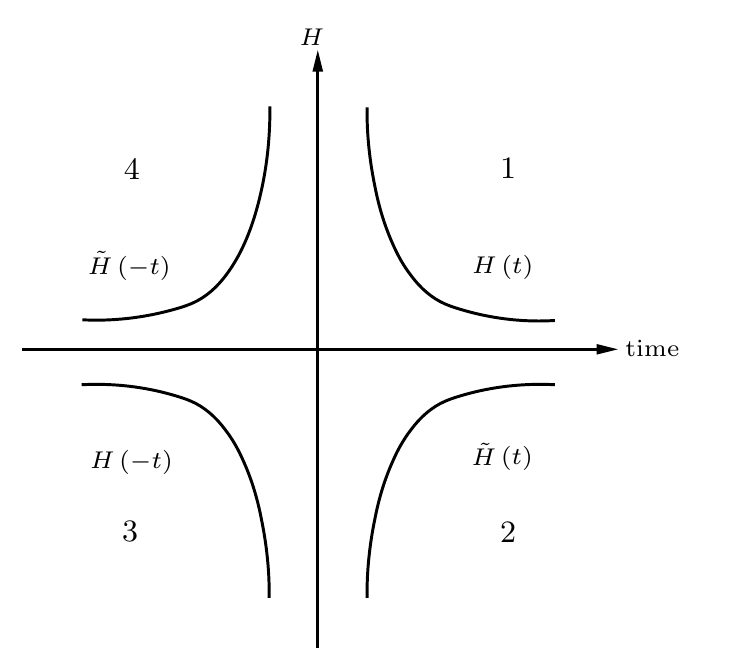}
\par\end{centering}

\caption{Hubble parameters in four solutions.}
\end{figure}

In order to group the solutions, a ``self-dual'' $\tilde
a\left(t\right)=a^{-1}\left(-t\right)$ is introduced. This duality pairs solution $1$ with $4$, an accelerated expansion followed by a decelerated expansion; solution $2$ with $3$, an accelerated contraction followed by a decelerated contraction. Each pair covers the whole spacetime except the singularity. It is natural to name the $t<0$ phase as the pre-big bang and the $t>0$ region as the post-big bang.

It is not surprising that, by including some matter sources or dilaton potentials, one can smooth the singularity to connect the pre- and post-big bangs, as illustrated in Figure $2$. These models are of great help to understand the physics in the region $t\sim 0$.

However, it should be noted that only one pair of solutions can be chosen to describe the evolution of the universe, between the two choices: $4\rightarrow 1$ or $3\rightarrow 2$. This observation makes it difficult for the standard string cosmology to embody extra dimensions and the widely accepted anisotropy in the early stage of the universe. We are going to show that DFT cosmology provides a natural way to solve these two problems in section $4$.

\begin{figure}[H]
\begin{centering}
\includegraphics{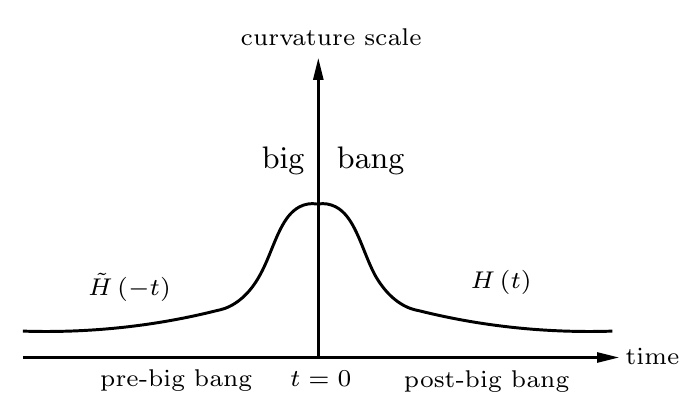}
\par\end{centering}

\caption{Pre-big bang, post-big bang and the avoided singularity.}
\end{figure}

\section{Equations of motion of double field theory}

The $O(D,D)$ invariance of the action enables us to construct infinitely many solutions from a single solution. However, general solutions of DFT, dependent both on $x$ and $\bar x$, are beyond the solution space of the low
energy effective action, and can only be calculated from the EOM of DFT. Therefore,  it is of importance to present the
EOM of DFT for general metrics and the FRW like metric.

\subsection{Equations of motion for general metrics}
Before we calculate the cosmological solutions, it is of help to
review the EOM of action (\ref{DFTS}) for general metrics.

\begin{eqnarray}
 &  & \frac{1}{8}\mathcal{H}^{MN}\partial_{M}\mathcal{H}^{KL}\partial_{N}
 \mathcal{H}_{KL}-\frac{1}{2}\mathcal{H}^{MN}\partial_{M}\mathcal{H}^{KL}
 \partial_{K}\mathcal{H}_{NL}\nonumber \\
 &  & -\partial_{M}\partial_{N}\mathcal{H}^{MN}-4\mathcal{H}^{MN}
 \partial_{M}d\partial_{N}d+4\partial_{M}\mathcal{H}^{MN}
 \partial_{N}d+4\mathcal{H}^{MN}\partial_{M}\partial_{N}d=0.
 \label{EOMd}
\end{eqnarray}

Next, we calculate the EOM of the graviton. Varying the action with
respect to the generalized metric $\mathcal{H}^{MN}$, we get

\begin{equation}
\delta_{\mathcal{H}}S=\int
dxd\tilde{x}e^{-2d}\delta\mathcal{H}^{MN}\mathcal{K}_{MN},
\label{VarySH}
\end{equation}

\noindent where

\begin{eqnarray}
\mathcal{K}_{MN} & \equiv & \frac{1}{8}\partial_{M}\mathcal{H}^{KL}\partial_{N}
\mathcal{H}_{KL}-\frac{1}{4}\left(\partial_{L}-2\partial_{L}d\right)\left(
\mathcal{H}^{LK}\partial_{K}\mathcal{H}_{MN}\right)+2\partial_{M}\partial_{N}d\nonumber \\
 &  & -\frac{1}{2}\partial_{\left(N\right.}\mathcal{H}^{KL}\partial_{L}
 \mathcal{H}_{\left.M\right)K}+\frac{1}{2}\left(\partial_{L}-2\partial_{L}
 d\right)\left(\mathcal{H}^{KL}\partial_{\left(N\right.}\mathcal{H}_{
 \left.M\right)K}+\mathcal{H}_{\quad\left(M\right.}^{K}\partial_{K}
 \mathcal{H}_{\quad\left.N\right)}^{L}\right).
\end{eqnarray}

\noindent However, $\mathcal{K}_{MN}=0$ is not the field equation,
since one needs to check the $O\left(D,D\right)$ symmetry of this
term. Following the definitions in \cite{Hohm:2010pp}, we take the
notations $\mathcal{H}\equiv\mathcal{H}^{\bullet\bullet}$ and
$\eta\equiv\mathcal{\eta}_{\bullet\bullet}$. $\mathcal{H}$ should
satisfy $\mathcal{H}\eta\mathcal{H}=\eta^{-1}$ to respect the
$O\left(D,D\right)$ symmetry. It is easy to see that the variation
of $\mathcal{H}$  satisfies the condition

\begin{equation}
\delta\mathcal{H}\eta\mathcal{H}+\mathcal{H}\eta\delta\mathcal{H}=0.
\end{equation}

\noindent It is convenient to define

\begin{equation}
S_{\quad N}^{M}\equiv\mathcal{H}_{\quad
N}^{M}=\eta^{MP}\mathcal{H}_{PN}=\mathcal{H}^{MP}\eta_{PN},\quad
S^{2}=1.
\end{equation}

\noindent Using the notation $S\equiv
S_{\quad\bullet}^{\bullet}=\mathcal{H}\eta$, one finds

\begin{equation}
\delta\mathcal{H}S^{t}+S\delta\mathcal{H}=0.
\end{equation}
Then since $S^{2}=1$, one has

\begin{equation}
\delta\mathcal{H}=-S\delta\mathcal{H}S^{t}.
\end{equation}

\noindent It can be rewritten as

\begin{equation}
\delta\mathcal{H}=\frac{1}{4}\left(1+S\right)\mathcal{M}\left(1-S^{t}\right)
+\frac{1}{4}\left(1-S\right)\mathcal{M}\left(1+S^{t}\right),
\end{equation}
where $\mathcal{M}$ is an arbitrary symmetric matrix to guarantee
the symmetry of $\delta\mathcal{H}$. Substituting it back into the
variation of the action (\ref{VarySH}), we find

\begin{equation}
\frac{1}{4}\left(1-S^{t}\right)\mathcal{K}\left(1+S\right)+\frac{1}{4}
\left(1+S^{t}\right)\mathcal{K}\left(1-S\right)=0.
\end{equation}

\noindent Inserting $O\left(D,D\right)$ indices $M$ and $N$ to
rewrite the equation above, the field equation of
$\mathcal{H}^{MN}$ is obtained

\begin{eqnarray}
\mathcal{R}_{MN} & = & \frac{1}{4}\left(\delta_{M}^{\quad P}-S_{\quad M}^{P}
\right)\mathcal{K}_{PQ}\left(\delta_{\quad N}^{Q}+S_{\quad N}^{Q}
\right)+\frac{1}{4}\left(\delta_{M}^{\quad P}+S_{\quad M}^{P}
\right)\mathcal{K}_{PQ}\left(\delta_{\quad N}^{Q}-S_{\quad N}^{Q}\right)\nonumber \\
 & = & \frac{1}{2}\mathcal{K}_{MN}-\frac{1}{2}S_{\quad M}^{P}
 \mathcal{K}_{PQ}S_{\quad N}^{Q}.
 \label{GenRicci}
\end{eqnarray}

\noindent From this equation, one finds the field equations of the
metric $g$ and the $b$ field. In summary, the EOM of the dilaton and
the graviton are

\begin{eqnarray}
\frac{1}{8}\mathcal{H}^{MN}\partial_{M}\mathcal{H}^{KL}\partial_{N}
\mathcal{H}_{KL}-\frac{1}{2}\mathcal{H}^{MN}\partial_{M}\mathcal{H}^{KL}
\partial_{K}\mathcal{H}_{NL}-\partial_{M}\partial_{N}\mathcal{H}^{MN}\nonumber \\
-4\mathcal{H}^{MN}\partial_{M}d\partial_{N}d+4\partial_{M}\mathcal{H}^{MN}
\partial_{N}d+4\mathcal{H}^{MN}\partial_{M}\partial_{N}d & = & 0,\label{EOMD} \\
\mathcal{K}_{MN}-S_{\quad M}^{P}\mathcal{K}_{PQ}S_{\quad N}^{Q} &
= & 0\label{EOMH}.
\end{eqnarray}

\subsection{Some notations and definitions}
We want to introduce some notations and  calculation rules before
performing the calculation. This helps us to track the double
coordinates when performing derivatives. For convenience, we use
block matrices to rewrite the vectors, the dual vectors and the
generalized metric

\begin{equation}
\partial_{M}=\left(\begin{array}{c}
\partial_{\mathbf{1}}\\
\partial_{\mathbf{2}}
\end{array}\right),\quad dX^{M}=\left(\begin{array}{c}
dX^{\mathbf{1}}\\
dX^{\mathbf{2}}
\end{array}\right),\quad\mathcal{H}_{MN}=\left(\begin{array}{cc}
\mathcal{H}_{\mathbf{1}\mathbf{1}} & \mathcal{H}_{\mathbf{1}\mathbf{2}}\\
\mathcal{H}_{\mathbf{2}\mathbf{1}} & \mathcal{H}_{\mathbf{2}\mathbf{2}}
\end{array}\right),\quad\mathcal{H}^{MN}=\left(\begin{array}{cc}
\mathcal{H}^{\mathbf{1}\mathbf{1}} & \mathcal{H}^{\mathbf{1}\mathbf{2}}\\
\mathcal{H}^{\mathbf{2}\mathbf{1}} & \mathcal{H}^{\mathbf{2}\mathbf{2}}
\end{array}\right).
\end{equation}

\noindent Here $\mathbf{1}$ represents the dual coordinate
$\tilde{x}_{i}$ and $\mathbf{2}$ corresponds to the usual
coordinate $x^{i}$. The components of the generalized metric are
divided into four parts. Each of them defines the metric of
spacetime and its dual as shown in \cite{Vaisman}

\begin{eqnarray}
\mathcal{H}_{\mathbf{1}\mathbf{1}}\left(\frac{\partial}{\partial\tilde{x}_{i}},
\frac{\partial}{\partial\tilde{x}_{j}}\right)=g^{ij}, &  & \mathcal{H}_{\mathbf{1}
\mathbf{2}}\left(\frac{\partial}{\partial\tilde{x}_{i}},\frac{\partial}
{\partial x^{j}}\right)=-g^{ik}b_{kj},\nonumber \\
\mathcal{H}_{\mathbf{2}\mathbf{1}}\left(\frac{\partial}{\partial
x^{i}},
\frac{\partial}{\partial\tilde{x}_{j}}\right)=b_{ik}g^{kj}, &  &
\mathcal{H}_{\mathbf{\mathbf{2}}\mathbf{2}}\left(\frac{\partial}
{\partial x^{i}},\frac{\partial}{\partial
x^{j}}\right)=g_{ij}-b_{ik}g^{kl}b_{lj}.
\end{eqnarray}

\noindent The generalized line element is

\begin{eqnarray}
dS^{2} & = & \mathcal{H}_{\mathbf{1}\mathbf{1}}dX^{\mathbf{1}}dX^{\mathbf{1}}
+\mathcal{H}_{\mathbf{1}\mathbf{2}}dX^{\mathbf{1}}dX^{\mathbf{2}}
+\mathcal{H}_{\mathbf{2}\mathbf{1}}dX^{\mathbf{2}}dX^{\mathbf{1}}
+\mathcal{H}_{\mathbf{2}\mathbf{2}}dX^{\mathbf{2}}dX^{\mathbf{2}},\nonumber \\
 & = & g^{ij}d\tilde{x}_{i}d\tilde{x}_{j}-g^{ik}b_{kj}d\tilde{x}_{i}dx^{j}+
 b_{ik}g^{kj}dx^{i}d\tilde{x}_{j}+\left(g_{ij}-b_{ik}g^{kl}b_{lj}\right)dx^{i}dx^{j}.
\end{eqnarray}

\noindent In this paper, we consider the simplest situation with
$b_{ij}=0$. The line element is simplified to

\begin{equation}
dS^{2}=\mathcal{H}_{\mathbf{1}\mathbf{1}}dX^{\mathbf{1}}dX^{\mathbf{1}}+
\mathcal{H}_{\mathbf{2}\mathbf{2}}dX^{\mathbf{2}}dX^{\mathbf{2}}=
\tilde{g}^{ij}d\tilde{x}_{i}d\tilde{x}_{j}+g_{ij}dx^{i}dx^{j}.
\end{equation}

\noindent  To exhibit the contraction of the $O\left(D,D\right)$
indices, we introduce  extra indices
$\mathcal{H}^{\mathbf{1}\left(i\right)\mathbf{1}\left(j\right)}$
to denote elements of block matrices. Therefore, the generalized
metric is rewritten with the extra indices

\begin{equation}
\mathcal{H}_{M\left(i\right)N\left(j\right)}=\left(\begin{array}{cc}
\mathcal{H}_{\mathbf{1}\left(i\right)\mathbf{1}\left(j\right)} & 0\\
0 & \mathcal{H}_{\mathbf{2}\left(i\right)\mathbf{2}\left(j\right)}
\end{array}\right)
= \left(%
\begin{array}{cc}
  \tilde g^{ij} & 0 \\
  0 & g_{ij} \\
\end{array}%
\right), \label{DefineComp}
\end{equation}

\noindent where $b_{ij}=0$ is assumed. Now, there exist two sets
of indices: the block matrix notations $M$,
$N=\mathbf{1},\mathbf{2}$ and the indices $i$, $j=1,2,\cdots D$ of
the components of block matrices. A contraction of $M$, $N$ is
given by

\begin{equation}
\mathcal{H}^{MN}\partial_{M}d\partial_{N}d=\mathcal{H}^{\mathbf{1}
\mathbf{1}}\partial_{\mathbf{1}}d\partial_{\mathbf{1}}d+
\mathcal{H}^{\mathbf{2}\mathbf{2}}\partial_{\mathbf{2}}d
\partial_{\mathbf{2}}d,
\end{equation}

\noindent and a contraction of $i$, $j$ takes the form

\begin{eqnarray}
\mathcal{H}^{M\left(i\right)N\left(j\right)}\partial_{M\left(i\right)}
d\partial_{N\left(j\right)}d & = & \mathcal{H}^{\mathbf{1}
\left(i\right)\mathbf{1}\left(j\right)}\partial_{\mathbf{1}
\left(i\right)}d\partial_{\mathbf{1}\left(j\right)}d+\mathcal{H}^{
\mathbf{2}\left(i\right)\mathbf{2}\left(j\right)}\partial_{
\mathbf{2}\left(i\right)}d\partial_{\mathbf{2}\left(j\right)}d\nonumber \\
 & = & \tilde{g}_{ij}\tilde{\partial}^{i}d\tilde{\partial}^{j}
 d+g^{ij}\partial_{i}d\partial_{j}d.
\end{eqnarray}

\noindent On the right hand side of (\ref{DefineComp}), we
introduce a barred $\tilde g^{ij}=g^{ij}$. The purpose of this
notation is to help us calculate the derivatives of the metric,
since there are two operators $\tilde{\partial}^{i}$ and
$\partial_{i}$. When $\partial_{\mathbf{1}}$ acts on the metric,
we use the barred $\tilde g$; when $\partial_{\mathbf{2}}$ acts on
the metric, we use the unbarred $g$. We give some illustrations

\begin{eqnarray}
\partial_{\mathbf{1}\left(k\right)}\mathcal{H}^{\mathbf{2}\left(i\right)\mathbf{2}\left(j\right)}
& = & \tilde{\partial}^{k}g^{ij}=\tilde{\partial}^{k}\tilde{g}^{ij},\nonumber \\
\partial_{\mathbf{2}\left(k\right)}\mathcal{H}^{\mathbf{1}\left(i\right)
\mathbf{1}\left(j\right)} & = & \partial_{k}\tilde{g}_{ij}=\partial_{k}g_{ij},\nonumber \\
\partial_{\mathbf{1}\left(p\right)}\mathcal{H}^{\mathbf{1}\left(i\right)\mathbf{1}
\left(j\right)}\partial_{\mathbf{2}\left(q\right)}\mathcal{H}_{\mathbf{1}
\left(i\right)\mathbf{1}\left(j\right)} & = & \tilde{\partial}^{p}
\tilde{g}_{ij}\partial_{q}\tilde{g}^{ij}=\tilde{\partial}^{p}\tilde{g}_{ij}\partial_{q}g^{ij},
\end{eqnarray}

\noindent and
\begin{equation}
\tilde{g}_{ij}\tilde{\partial}^{i}=\tilde{\partial}_{j},\quad
g^{ij}\partial_{i}=\partial^{j},\quad\tilde{g}_{ij}g^{jk}=
g_{ij}\tilde{g}^{jk}=\delta_{i}^{k}.
\end{equation}

\noindent Two other useful relations are

\begin{eqnarray}
\partial_{\mathbf{2}\left(k\right)}\mathcal{H}_{\quad\mathbf{2}\left(j\right)}^{
\mathbf{1}\left(i\right)} & = & \partial_{k}\tilde{g}_{ij}=\partial_{k}g_{ij},\nonumber \\
\partial_{\mathbf{1}\left(k\right)}\mathcal{H}_{\quad\mathbf{1}\left(j\right)}^{
\mathbf{2}\left(i\right)} & = &
\tilde{\partial}^{k}g^{ij}=\tilde{\partial}^{k}\tilde{g}^{ij}.
\end{eqnarray}

\subsection{Equations of motion for FRW like metric}

We first consider the equation of motion of the dilaton (\ref{EOMD})
and expand it in components. For reference, we rewrite it

\begin{eqnarray}
 &  & \frac{1}{8}\mathcal{H}^{MN}\partial_{M}\mathcal{H}^{KL}\partial_{N}
 \mathcal{H}_{KL}-\frac{1}{2}\mathcal{H}^{MN}\partial_{M}\mathcal{H}^{KL}
 \partial_{K}\mathcal{H}_{NL}\nonumber \\
 &  & -\partial_{M}\partial_{N}\mathcal{H}^{MN}-4\mathcal{H}^{MN}
 \partial_{M}d\partial_{N}d+4\partial_{M}\mathcal{H}^{MN}\partial_{N}d
 +4\mathcal{H}^{MN}\partial_{M}\partial_{N}d=0.
 \label{EOMD_1}
\end{eqnarray}

\noindent Substituting the matrix components defined in
(\ref{DefineComp}), after some simplifications, we get

\begin{eqnarray}
 &  & \frac{1}{8}\tilde{g}_{ij}\tilde{\partial}^{i}\tilde{g}_{kl}\tilde
 {\partial}^{j}\tilde{g}^{kl}+\frac{1}{8}
 \tilde{g}_{ij}\tilde{\partial}^{i}g^{kl}\tilde{\partial}^{j}g_{kl}\nonumber \\
 &  & +\frac{1}{8}g^{ij}\partial_{i}\tilde{g}_{kl}\partial_{j}\tilde{g}^{kl}+\frac{1}
 {8}g^{ij}\partial_{i}g^{kl}\partial_{j}g_{kl}\nonumber \\
 &  & -\frac{1}{2}\tilde{g}_{ij}\tilde{\partial}^{i}\tilde{g}_{kl}\tilde{\partial}^{k}
 \tilde{g}^{jl}-\frac{1}{2}g^{ij}\partial_{i}g^{kl}\partial_{k}g_{jl}\nonumber \\
 &  & -\tilde{\partial}^{i}\tilde{\partial}^{j}\tilde{g}_{ij}-\partial_{i}
 \partial_{j}g^{ij}-4\tilde{g}_{ij}\tilde{\partial}^{i}d\tilde{\partial}^{j}
 d-4g^{ij}\partial_{i}d\partial_{j}d\nonumber \\
 &  &
 +4\tilde{\partial}^{i}\tilde{g}_{ij}\tilde{\partial}^{j}d+4\partial_{i}
 g^{ij}\partial_{j}d+4\tilde{g}_{ij}\tilde{\partial}^{i}\tilde{\partial}^{j}
 d+4g^{ij}\partial_{i}\partial_{j}d=0.
\end{eqnarray}

\noindent In double field theory, we know that the relationship of
the three dilatons is

\begin{equation}
e^{-2d}=\sqrt{g}e^{-2\phi}= \sqrt{\tilde g}e^{-2\tilde\phi}
\end{equation}

\noindent When applied to the metric ansatz (\ref{MetricAnsatz}),
we find

\begin{equation}
\phi=\tilde{\phi}+\left(D-1\right)\ln a, \hspace{5mm} d = \phi
-\frac{D-1}{2}\ln a = \tilde\phi - \frac{D-1}{2}\ln \tilde a
\end{equation}

\noindent The first equation is precisely the so-called
``scale-factor duality'' transformations  (\ref{SFDuality})  in
string cosmology. The second equation proves our claim that the
``shifted dilaton'' in string cosmology, defined in
(\ref{ShiftedD}), is $2d$. Then, with the metric ansatz
(\ref{MetricAnsatz}), the EOM of the dilaton is

\begin{eqnarray}
-\left(D-1\right)\frac{\dot{\tilde{a}}^{2}}{\tilde{a}^{2}}+\left(\left(D-1\right)\frac{\dot{\tilde{a}}}
{\tilde{a}}\right)^{2}-4\left(D-1\right)\frac{\dot{\tilde{a}}}{\tilde{a}}\dot{\tilde{\phi}}
+4\dot{\tilde{\phi}}^{2}+2\left(D-1\right)\frac{\ddot{\tilde{a}}}{\tilde{a}}-4\ddot{\tilde{\phi}}\nonumber \\
-\left(D-1\right)\frac{\dot{a}^{2}}{a^{2}}+\left(\left(D-1\right)\frac{\dot{a}}{a}\right)^{2}
-4\left(D-1\right)\frac{\dot{a}}{a}\dot{\phi}+4\dot{\phi}^{2}+2\left(D-1\right)\frac{\ddot{a}}{a}-4\ddot{\phi}
& = & 0,
\end{eqnarray}
where $\dot a= \frac{da}{dt}$ and $\dot{\tilde a}=\frac{d{\tilde
a}}{d\tilde t}$.  In order to compare the equations from string
cosmology, we used $\phi$ and $\tilde\phi$ but not $d$. To avoid confusion of $a$ and $\tilde
a$, we replace $\tilde a = a^{-1}$  in the EOM and define the
Hubble parameters
\begin{equation}
H=\frac{\partial_{t}a}{a},\qquad\tilde{H}=\frac{\partial_{\tilde{t}}a}{a}.
\end{equation}

\noindent Then the EOM of the dilaton for our metric ansatz
(\ref{MetricAnsatz}) is

\begin{eqnarray}
\left(\left(D-1\right)\tilde{H}\right)^{2}+4\left(D-1\right)\tilde{H}\dot{\tilde{\phi}}+4\dot{\tilde{\phi}}^{2}-2\left(D-1\right)\dot{\tilde{H}}+\left(D-1\right)\tilde{H}^{2}-4\ddot{\tilde{\phi}}\nonumber \\
+\left(\left(D-1\right)H\right)^{2}-4\left(D-1\right)H\dot{\phi}+4\dot{\phi}^{2}+2\left(D-1\right)\dot{H}+\left(D-1\right)H^{2}-4\ddot{\phi} & = & 0.\label{EOMDMetric}
\end{eqnarray}

\noindent The EOM of the graviton is given in (\ref{EOMH}). We
rewrite it for reference
\begin{equation}
\mathcal{R}_{MN} = \mathcal{K}_{MN}-S_{\quad
M}^{P}\mathcal{K}_{PQ}S_{\quad N}^{Q} = 0.
\end{equation}

\noindent Refer to the Appendix, we find there exist symmetries
between the components of the generalized Ricci tensor

\begin{eqnarray}
\mathcal{R}_{\mathbf{1}\left(p\right)\mathbf{1}\left(q\right)} &
\underleftrightarrow{g^{\bullet\bullet}\leftrightarrow g_{\bullet\bullet},
\quad\tilde{\partial}^{\bullet}\leftrightarrow\partial_{\bullet},\quad
\tilde{\phi}\leftrightarrow\phi} & \mathcal{R}_{\mathbf{2}\left(p\right)
\mathbf{2}\left(q\right)},\nonumber \\
\mathcal{R}_{\mathbf{1}\left(p\right)\mathbf{2}\left(q\right)} &
\underleftrightarrow{g^{\bullet\bullet}\leftrightarrow
g_{\bullet\bullet},\quad\tilde{\partial}^{\bullet}\leftrightarrow
\partial_{\bullet},\quad\tilde{\phi}\leftrightarrow\phi}
& \mathcal{R}_{\mathbf{2}\left(p\right)\mathbf{1}\left(q\right)}.
\end{eqnarray}
Also in the Appendix, we present the lengthy calculation process
and find the EOM of the graviton

\begin{eqnarray}
\mathcal{R}_{\mathbf{2}\left(t\right)\mathbf{2}\left(t\right)} & = & -\left(D-1\right)\frac{\ddot{a}}{a}+2\ddot{\phi}+\left(D-1\right)\frac{\ddot{\tilde{a}}}{\tilde{a}}-2\ddot{\tilde{\phi}}\nonumber \\
 & = & -\left(D-1\right)\left(\dot{H}+H^{2}\right)+2\ddot{\phi}+\left(D-1\right)\left(-\dot{\tilde{H}}+\tilde{H}^{2}\right)-2\ddot{\tilde{\phi}},\nonumber \\
\nonumber \\
\mathcal{R}_{\mathbf{2}\left(i\right)\mathbf{2}\left(i\right)} & = & \frac{1}{\tilde{a}^{2}}\left(\frac{\dot{\tilde{a}}^{2}}{\tilde{a}^{2}}-\frac{\ddot{\tilde{a}}}{\tilde{a}}-\left(D-1\right)\frac{\dot{\tilde{a}}^{2}}{\tilde{a}^{2}}+2\dot{\tilde{\phi}}\frac{\dot{\tilde{a}}}{\tilde{a}}\right)-a^{2}\left(\frac{\dot{a}^{2}}{a^{2}}-\frac{\ddot{a}}{a}-\left(D-1\right)\frac{\dot{a}^{2}}{a^{2}}+2\dot{\phi}\frac{\dot{a}}{a}\right)\nonumber \\
 & = & {a}^{2}\left(\dot{\tilde{H}}-\left(D-1\right)\tilde{H}^{2}-2\dot{\tilde{\phi}}\tilde{H}\right)-a^{2}\left(-\dot{H}-\left(D-1\right)H^{2}+2\dot{\phi}H\right),\nonumber \\
\nonumber \\
\mathcal{R}_{\mathbf{1}\left(t\right)\mathbf{2}\left(t\right)} & = & 0.
\end{eqnarray}
Including the EOM of the dilaton (\ref{EOMDMetric}), the set of
equations we need to solve is

\begin{eqnarray}
\left(\left(D-1\right)\tilde{H}\right)^{2}+4\left(D-1\right)\tilde{H}\dot{\tilde{\phi}}+4\dot{\tilde{\phi}}^{2}-2\left(D-1\right)\dot{\tilde{H}}+\left(D-1\right)\tilde{H}^{2}-4\ddot{\tilde{\phi}}\nonumber \\
+\left(\left(D-1\right)H\right)^{2}-4\left(D-1\right)H\dot{\phi}+4\dot{\phi}^{2}+2\left(D-1\right)\dot{H}+\left(D-1\right)H^{2}-4\ddot{\phi} & = & 0,\nonumber \\
-\left(D-1\right)\left(\dot{H}+H^{2}\right)+2\ddot{\phi}+\left(D-1\right)\left(-\dot{\tilde{H}}+\tilde{H}^{2}\right)-2\ddot{\tilde{\phi}} & = & 0,\nonumber \\
\left(\dot{\tilde{H}}-\left(D-1\right)\tilde{H}^{2}-2\dot{\tilde{\phi}}\tilde{H}\right)-\left(-\dot{H}-\left(D-1\right)H^{2}+2\dot{\phi}H\right) & = & 0.\label{FinalEOM}
\end{eqnarray}


\noindent Now we replace $\phi$ and $\tilde{\phi}$ by the $O\left(D,D\right)$ scalar dilaton $d$
\begin{equation}
\phi  =  d+\frac{1}{2}\left(D-1\right)\ln a,\hspace{5mm}
\tilde{\phi}  =  d-\frac{1}{2}\left(D-1\right)\ln a.
\end{equation}

\noindent Eventually, the EOM are

\vspace{1cm}

\fbox{\begin{minipage}[t]{1\linewidth}
\begin{eqnarray}
\left(4\partial_{\tilde{t}}\partial_{\tilde{t}}d-4\left(\partial_{\tilde{t}}d\right)^{2}-\left(D-1\right)\tilde{H}^{2}\right)+\left(4\partial_{t}\partial_{t}d-4\left(\partial_{t}d\right)^{2}-\left(D-1\right)H^{2}\right) & = & 0,\nonumber \\
\left(-\left(D-1\right)H^{2}+2\partial_{t}\partial_{t}d\right)-\left(-\left(D-1\right)\tilde{H}^{2}+2\partial_{\tilde{t}}\partial_{\tilde{t}}d\right) & = & 0,\nonumber \\
\left(\dot{\tilde{H}}-2\tilde{H}\partial_{\tilde{t}}d\right)+\left(\dot{H}-2H\partial_{t}d\right) & = & 0.
\label{eq: final EOM}
\end{eqnarray}
\end{minipage}}

\vspace{1cm}

\noindent Clearly, the barred part and unbarred parts are
identical, being the same as the EOM (\ref{EOMCom}) in string
cosmology. This indicates that if $a(t,\tilde t)$ is a solution, an $O\left(D,D\right)$ rotation of $a(t, \tilde t)$ is also a solution, as one can expects from the explicit $O\left(D,D\right)$ invariance in the action. One can easily check that

\begin{eqnarray}
a_{\pm}\left(\tilde{t},t\right)=\left|\frac{t}{\tilde{t}}\right|^{\pm1/\sqrt{D-1}}, & d\left(t\right)=-\frac{1}{2}\ln|t\,\tilde{t}|,\nonumber \\
a_{\pm}\left(\tilde{t},t\right)=\left|t\,\tilde{t}\right|^{\pm1/\sqrt{D-1}}, & d\left(t\right)=-\frac{1}{2}\ln|t\,\tilde{t}|,\label{eq:violation solution}
\end{eqnarray}
are solutions of the EOM. These solutions are beyond the solution space of string cosmology and only exist in DFT cosmology, though they violate the constraints. However, we have no clue how to interpret these solutions, especially since there exist double times.

\section{Cosmological solutions}

In this section, we address solutions  dependent on only one set of coordinates, one time-like coordinate in particular. Then, the action takes a form
\begin{equation}
S=\int d^D x d^D\tilde x \mathcal{L}(x) = T \int d^D x d^{D-1}\tilde x \mathcal{L}(x),
\end{equation}
where $T=\int d\tilde t$. Therefore, the metric includes only one timelike direction.  we will calculate the cosmological solutions with
vanishing and non-vanishing dilation potentials respectively.

\subsection{Solutions with $V(d)=0$}

\begin{figure}[H]
\begin{centering}
\includegraphics[scale=0.7]{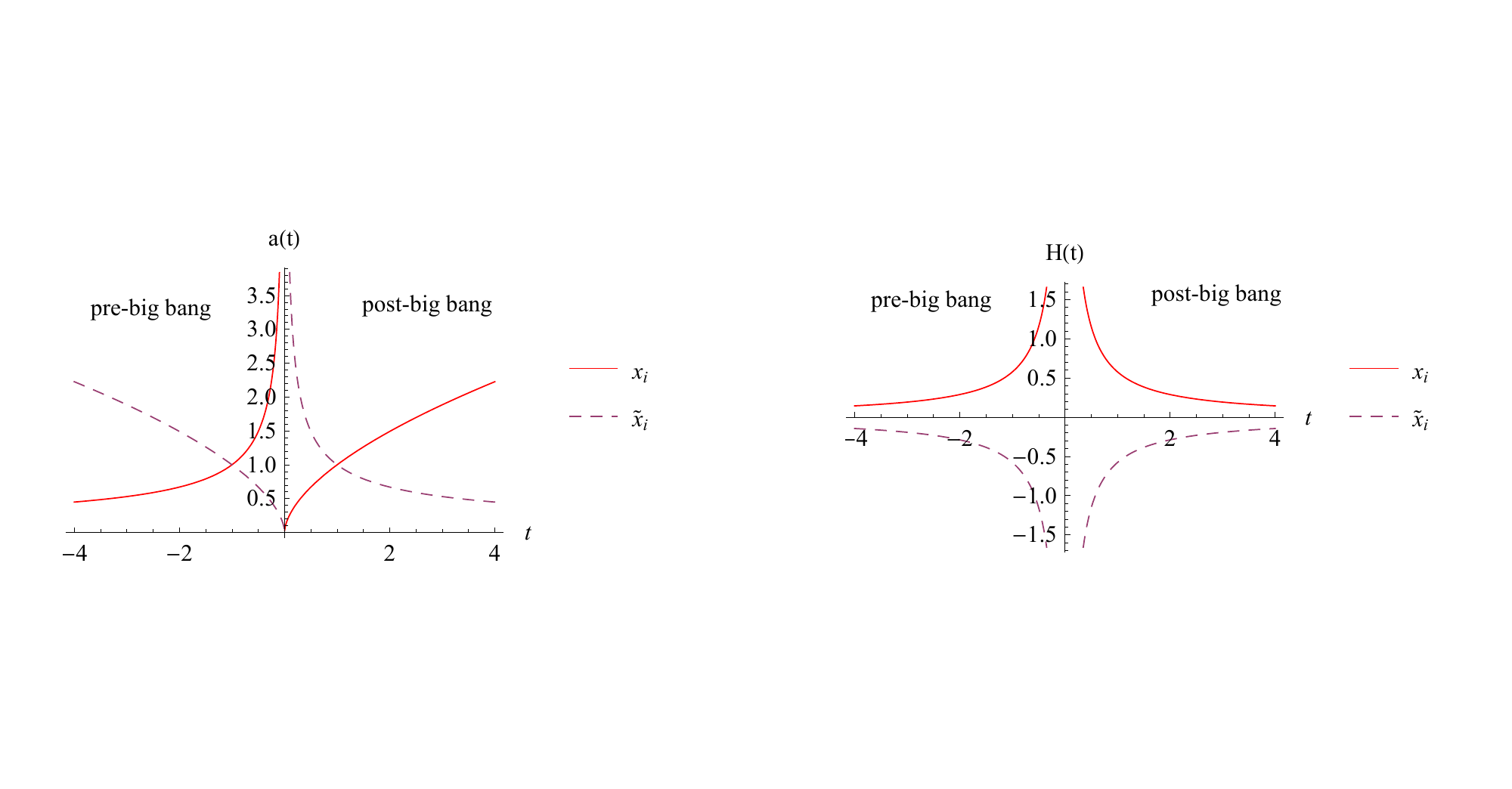}
\par\end{centering}
\caption{The evolutions for $V(d)=0$. The left hand side figure describes the scale factors and the right hand side one represents the Hubble parameters. The solid line stands for the evolutions of the ordinary coordinates $x_i$ and the dashed line is  the evolutions of the dual coordinates $\tilde x_i$.}
\label{fig: V=0 scale}
\end{figure}

The solutions are easily obtained from these of string cosmology
\begin{equation}
a(t)= |t|^{1/\sqrt{D-1}}, \hspace{5mm} d(t) = -\frac{1}{2}\ln|t|,
\end{equation}
where we set the initial time to the unity. In this work, we choose $D=4$. The metrics (\ref{MetricAnsatz}) become

\begin{align}
dS_{pre}^{2} & =-dt^{2}+a\left(-t\right)^{-2}\left(dx_{2}^{2}+dx_{3}^{2}+dx_{4}^{2}\right)+a\left(-t\right)^{2}\left(d\tilde{x}_{2}^{2}+d\tilde{x}_{3}^{2}+d\tilde{x}_{4}^{2}\right),\qquad t<0,\nonumber \\
\nonumber \\
dS_{post}^{2} & =-dt^{2}+a\left(t\right)^{2}\left(dx_{2}^{2}+dx_{3}^{2}+dx_{4}^{2}\right)+a\left(t\right)^{-2}\left(d\tilde{x}_{2}^{2}+d\tilde{x}_{3}^{2}+d\tilde{x}_{4}^{2}\right),\qquad t>0.
\end{align}
where, we have selected the ``self-dual'' evolutions: an accelerated expansion followed by a decelerated expansion of $x$, an accelerated contraction evolving to a decelerated contraction of $\tilde x$, as illustrated in FIG. (\ref{fig: V=0 scale})

New physics already show up evidently in the solutions. However, we will put off the discussions to the $V(d)\not=0$ scenario. Since the singularity can be smoothed out by a nontrivial dilaton potential, we can see the novel features more clearly. Moreover, the reason to choose the self-dual evolution will be justified.

\subsection{Solutions with $V(d)\neq 0$}

\noindent To remove the singularity, we introduce a dilaton potential

\begin{equation}
V\left(d\right)=V_{0}e^{8d}.
\end{equation}

\noindent where $V_{0}>0$. This non-local potential represents the backreactions of higher loop corrections \cite{Gasperini:2003pb}. This potential certainly respect the $O(D,D)$ symmetry. In physics, we alway give symmetries the highest priority. Therefore, its presence in the action is well justified. Anyhow, the primary physics are not affected by the potential.  It is easy to figure out that the physically relevant solution is

\begin{equation}
a\left(t\right)=a_{0}\left[\frac{t}{t_{0}}+\left(1+\frac{t^{2}}{t_{0}^{2}}\right)^{\frac{1}{2}}\right]^{\frac{1}{\sqrt{D-1}}},\qquad d\left(t\right)=-\frac{1}{4}\ln\left[\sqrt{V_{0}}t_{0}\left(1+\frac{t^{2}}{t_{0}^{2}}\right)\right].
\end{equation}

%

\begin{figure}[H]
\begin{centering}
\includegraphics[scale=0.7]{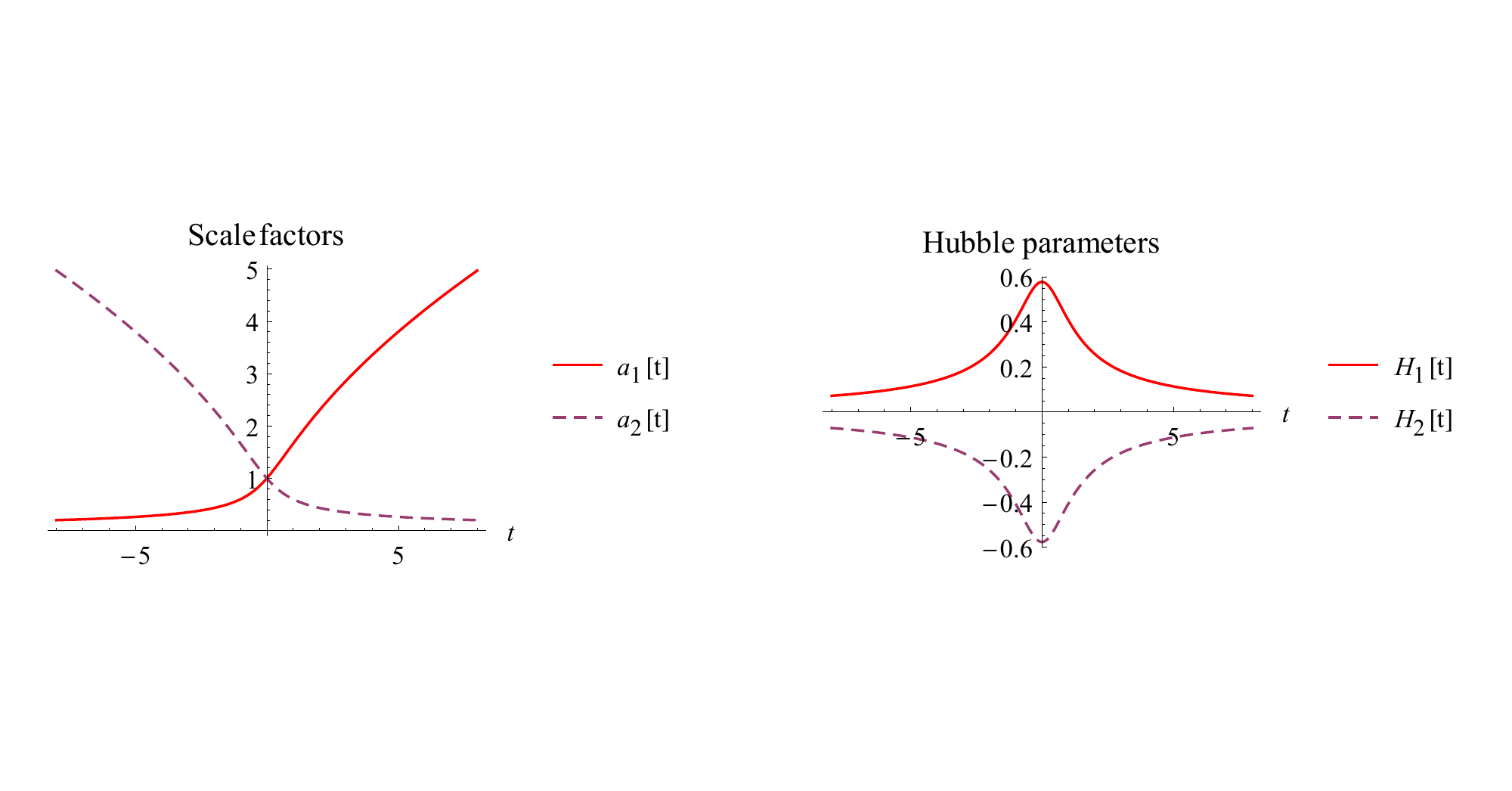}
\par\end{centering}
\caption{The evolutions for $V(d)=V_0 e^{8d}$. The left hand side figure describes the scale factors and the right hand side one represents the Hubble parameters. The solid line stands for the evolutions of the ordinary coordinates $x_i$ and the dashed line is  the evolutions of the dual coordinates $\tilde x_i$.}
\label{fig: V nonezero scale}
\end{figure}

\noindent We set $a_0=t_{0}=V_{0}=1$ and $D=4$ for simplicity. The unified line element of DFT cosmology is

\begin{equation}
dS^{2}=-dt^{2}+a_{1}\left(t\right)^{2}\left(dx_{2}^{2}+dx_{3}^{2}+dx_{4}^{2}\right)+a_{2}\left(t\right)^{2}\left(d\tilde{x}_{2}^{2}+d\tilde{x}_{3}^{2}+d\tilde{x}_{4}^{2}\right).
\end{equation}

\noindent where $a_{1}\left(t\right)=a\left(t\right)$ and $a_{2}\left(t\right)=a\left(t\right)^{-1}$. This solution is unique in the sense that $x$ and $\tilde x$ are completely equivalent.
The evolutions of the scale factors and Hubble parameters are illustrated in FIG. (\ref{fig: V nonezero scale})

\noindent

\noindent \underline{In the pre-big bang region ($t\rightarrow-\infty$)}

\noindent
\begin{equation}
a_{1}\left(t\right)\sim\left(-t\right)^{-\frac{1}{\sqrt{3}}},\qquad a_{2}\left(t\right)\sim\left(-t\right)^{\frac{1}{\sqrt{3}}}.
\end{equation}
The ordinary spatial dimensions $x_i$'s are hidden, while the dual spatial dimensions $\tilde x_i$'s are visible.

\hspace{3ex}

\noindent \underline{In the post-big bang region ($t\rightarrow\infty$)}

\begin{equation}
a_{1}\left(t\right)\sim t^{\frac{1}{\sqrt{3}}},\qquad a_{2}\left(t\right)\sim t^{-\frac{1}{\sqrt{3}}}.
\end{equation}
It is obvious that $x_i$'s expand to be visible and effectively isotropic. Dual coordinates $\tilde x_i$'s contract to extra dimensions.
\hspace{3ex}

\noindent \underline{In the big bang region ($t\sim 0$)}

\begin{eqnarray}
a_{1}\left(t\right)&=&\kappa_0 + \kappa_1 \Big(t- \frac{1}{\sqrt 2}\Big) - \kappa_3 \Big(t- \frac{1}{\sqrt 2}\Big)^3 + \mathcal{O}\Big(t- \frac{1}{\sqrt 2}\Big)^4,\nonumber\\
a_{2}\left(t\right)&=&\kappa_0 - \kappa_1 \Big(t+ \frac{1}{\sqrt 2}\Big) + \kappa_3 \Big(t+ \frac{1}{\sqrt 2}\Big)^3 + \mathcal{O}\Big(t+ \frac{1}{\sqrt 2}\Big)^4,
\end{eqnarray}
where $\kappa_0, \kappa_1$ and $\kappa_3$ are positive numbers. One can easily see that the ordinary spatial dimensions $x_i$'s inflate all the way from $t\to -\infty$ to  $t=1/\sqrt 2$. After that, $x_i$'s start a decelerated expansion. On the other hand, all $\tilde x_i$'s first experience an accelerated contraction until $t=-1/\sqrt 2$, followed by a decelerated contraction to $t\to \infty$. This picture confirms our ``self-dual'' choice in the $V(d)=0$ scenario.

\noindent We see that the absence of the singularity makes the physical picture much clearer.  Two novel physical features come out immediately
\begin{enumerate}
\item Extra dimensions intrinsically show up in the asymptotic regions ($|t|\to \infty$).
\item Without any fine-tunning of parameters, we have a natural evolution from a visibly isotropic pre-big bang to an evidently anisotropic big bang and then again to  an isotropic universe at present time.
\end{enumerate}

Bear in mind that string cosmology can have only one of the evolutions, always expanding or contracting. Therefore, in order to have extra dimensions as ingredients of the models, one has to add them by hand, which introduces lot of arbitrariness. While, in DFT cosmology, thanks to the $O(D,D)$ symmetry, the existence of extra dimensions arises without any pre-assumptions. This fact again asserts the importance of symmetries in physics.

It is of interest to compare DFT cosmology with compactified Kaluza-Klein
gravity in the post-big bang region, since they have similar metrics. The higher dimensional Kaluza-Klein
gravity takes a form \cite{Overduin:1998pn}

\noindent
\begin{equation}
ds_{_{KK}}^{2}=-dt^{2}+t^{\alpha}\left(dx_{2}^{2}+dx_{3}^{2}+dx_{4}^{2}\right)+t^{-\alpha}dy^{2},
\end{equation}
where $\alpha$ is constant and $y$ represents an extra dimension.
The five dimensional Brans\textendash{}Dicke Theory also has the same
line element \cite{Bahrehbakhsh:2010cx}. It is immediately to see that DFT cosmology possesses all the properties of these theory, but without man-made setups.

Furthermore, our solutions show that, when traced back along the evolution, our current universe was totally hidden in the pre-big bang. While the visible dimensions  $\left(d\tilde{x}_{2}^{2},d\tilde{x}_{3}^{2},d\tilde{x}_{4}^{2}\right)$ in the pre-big bang become extra dimensions at present time. Two groups of spaces have interactions around the big bang region. Therefore,  exploring the extra dimensions could reveal the existence  and information of the pre-big bang.

In \cite{Chodos:1980}, the authors showed that the shear of the contracting dimensions causes a decelerated expansion of other dimensions.
In \cite{Levin:1994yw}, Levin demonstrated that if the dimensionality of the contracting extra dimensions is larger than $1$,  a kinetic inflation purely driven by the contraction of extra dimensions is also possible. However, the number of expanding/contracting dimensions has to be specified as initial conditions. These models further suffer graceful exit
and isotropy problem. Remarkably, as we see in DFT cosmology, both scenarios are automatically achieved without any
initial data. In the domain $-\infty <t<1/\sqrt 2$, the contraction of $\tilde x$ inflates $x$ and in the domain $1/\sqrt 2< t< \infty$, the contraction of $\tilde x$ makes $x$ expanding in an decelerated pace.

\section{Conclusion and Discussions}

In this paper, we calculated the cosmological solutions of double
field theory. We set the anti-symmetrical Kalb-Ramond field
vanishing for simplicity. When taking the FRW like metric ansatz, we
demonstrated that, the scale factor dual dilatons,
$\phi=\tilde{\phi}+\left(D-1\right)\ln a$, in string cosmology are
exactly the diffeomorphic and dual diffeomorphic dilatons in
double field theory. The ``shifted dilaton'' in string cosmology
is really the $O(D,D)$ scalar dilaton in double field theory with
$2d=\tilde{\psi}=\psi$.

We found two cosmological solutions, with and without an $O(D,D)$ invariant dilaton potential. In the $V(d)=0$ scenario, solutions
have two distinct metrics, representing the pre- and post-big bangs, respectively. Each of them
unifies contracting and expanding dimensions. As $t\rightarrow 0$, all solutions approach the
big bang singularity. To understand the physics around the singularity more clearly, we make use of an $O(D,D)$ invariant dilaton potential, $V(d) = V_0 e^{8d}$, $V_0>0$,
which does not  affect the main conclusions. Not only does this potential preserve the symmetry, it has physical origin from the backreactions of higher loop corrections.
With this potential, the big bang singularity disappears.  We thus have a single line element which unifies originally disconnected pre- and post-big bang metrics.
The visible pre-big bang dimensions contract to  invisible extra
dimensions. While extra dimensions in the pre-big bang expand all the way to the visible dimensions of the present universe. Due to this observation, one can expect that detection of extra dimension will reveal information of the pre-big bang.

In addition, we showed that the contraction of the dual dimensions causes both an inflation and a decelerated expansion of the ordinary dimensions in different time domains. The advantage of DFT cosmology is that no initial conditions are needed to set up such scenarios.

The solutions we have obtained are special ones of EOM
(\ref{eq: final EOM}). We also presented some constraint violating solutions. However, the physical implications of these solutions are unclear.
Though it looks not easy to figure out other more nontrivial
solutions, it would be of interest if one can find some.

Looking for other solutions of the generalized action, say black
holes, is of interest. However, the physical interpretations are
blurry. One has to be careful to deal with the gauge constraint
and identify the parameters in the solutions.

In the formulation of DFT, the weak and strong constraints are
sufficient but not necessary for the consistency of the theory. It is possible
to relax these constraints in some scenarios of flux
compactification and dimensional reduction \cite{Hohm:2011cp,
Aldazabal:2011nj, Geissbuhler:2011mx, Grana:2012rr,
Andriot:2012wx, Andriot:2012an, Geissbuhler:2013uka}. It is of
interest to investigate the relevance of our solutions to these
compactifications in the future works.

\bigskip
\noindent {\bf Acknowledgements} We are indebted to B. Zwiebach
for useful conversations. We would like to acknowledge illuminating discussions with T. Li, J. Lu, Z. Sun and P. Wang. This work is supported in part by the NSFC (Grant No. 11175039 and 11375121) and SiChuan Province Science Foundation for Youths (Grant No. 2012JQ0039). H. Y. is grateful to the hospitality of the Institute of Theoretical Physics, Chinese Academy of Sciences where part of this work is done..
\bigskip

\appendix
\renewcommand{\theequation}{\Alph{section}.\arabic{equation}}
\section{Equations of motion of the gravitons}
\setcounter{equation}{0}

\noindent Recall the generalized Ricci tensor (\ref{GenRicci}),

\noindent
\begin{equation}
\mathcal{R}_{MN}=\mathcal{K}_{MN}-S_{\quad
M}^{P}\mathcal{K}_{PQ}S_{\quad N}^{Q}=0.
\end{equation}

\noindent To simplify the calculations, we sperate
$\mathcal{K}_{MN}$ into two parts

\noindent
\begin{equation}
\mathcal{K}_{MN}=\star\mathcal{K}_{MN}+*\mathcal{K}_{MN},
\end{equation}

\noindent where

\noindent
\begin{eqnarray}
\star\mathcal{K}_{MN} & \equiv &
\frac{1}{8}\partial_{M}\mathcal{H}^{KL}\partial_{N}
\mathcal{H}_{KL}-\frac{1}{4}\partial_{L}\left(\mathcal{H}^{LK}\partial_{K}
\mathcal{H}_{MN}\right)\nonumber \\
 &  & -\frac{1}{2}\partial_{\left(N\right.}\mathcal{H}^{KL}\partial_{L}\mathcal{H}_{
 \left.M\right)K}+\frac{1}{2}\partial_{L}\left(\mathcal{H}^{KL}\partial_{\left(N\right.}
 \mathcal{H}_{\left.M\right)K}+\mathcal{H}_{\quad\left(M\right.}^{K}\partial_{K}
 \mathcal{H}_{\quad\left.N\right)}^{L}\right),
\end{eqnarray}

\noindent and

\noindent
\begin{equation}
*\mathcal{K}_{MN}\equiv\frac{1}{2}\partial_{L}d\left(\mathcal{H}^{LK}\partial_{K}
\mathcal{H}_{MN}\right)+2\partial_{M}\partial_{N}d-\partial_{L}d
\left(\mathcal{H}^{KL}\partial_{\left(N\right.}\mathcal{H}_{\left.M\right)K}+
\mathcal{H}_{\quad\left(M\right.}^{K}\partial_{K}
\mathcal{H}_{\quad\left.N\right)}^{L}\right).
\end{equation}

\noindent Therefore, $\mathcal{R}_{MN}=0$ can be put into two
parts

\noindent
\begin{equation}
\mathcal{R}_{MN}=\star\mathcal{R}_{MN}+*\mathcal{R}_{MN}=0,
\end{equation}

\noindent with

\noindent
\begin{eqnarray}
\star\mathcal{R}_{MN} & = & \star\mathcal{K}_{MN}-S_{\quad M}^{P}\star\mathcal{K}_{PQ}S_{\quad N}^{Q},\nonumber \\
*\mathcal{R}_{MN} & = & *\mathcal{K}_{MN}-S_{\quad
M}^{P}*\mathcal{K}_{PQ}S_{\quad N}^{Q}.
\end{eqnarray}

\section*{\noindent \underline{Calculation of $\star\mathcal{R}_{MN}$}}

\noindent
\begin{eqnarray}
\star\mathcal{R}_{MN} & = & \star\mathcal{K}_{MN}-S_{\quad M}^{P}\star\mathcal{K}_{PQ}S_{\quad N}^{Q}\nonumber \\
 & = & \frac{1}{8}\partial_{M}\tilde{g}_{ij}\partial_{N}\tilde{g}^{ij}+\frac{1}{8}\partial_{M}g^{ij}\partial_{N}g_{ij}\nonumber \\
 &  & -\frac{1}{4}\tilde{\partial}^{i}\left(\tilde{g}_{ij}\tilde{\partial}^{j}\mathcal{H}_{MN}\right)-\frac{1}{4}\partial_{i}\left(g^{ij}\partial_{j}\mathcal{H}_{MN}\right)\nonumber \\
 &  & -\frac{1}{4}\partial_{N}\tilde{g}_{ij}\tilde{\partial}^{j}\mathcal{H}_{M\mathbf{1}\left(i\right)}-\frac{1}{4}\partial_{M}\tilde{g}_{ij}\tilde{\partial}^{j}\mathcal{H}_{N\mathbf{1}\left(i\right)}\nonumber \\
 &  & -\frac{1}{4}\partial_{N}g^{ij}\partial_{j}\mathcal{H}_{M\mathbf{2}\left(i\right)}-\frac{1}{4}\partial_{M}g^{ij}\partial_{j}\mathcal{H}_{N\mathbf{2}\left(i\right)}\nonumber \\
 &  & +\frac{1}{4}\tilde{\partial}^{i}\left(\tilde{g}_{ji}\partial_{N}\mathcal{H}_{M\mathbf{1}\left(j\right)}\right)+\frac{1}{4}\tilde{\partial}^{i}\left(\tilde{g}_{ji}\partial_{M}\mathcal{H}_{N\mathbf{1}\left(j\right)}\right)\nonumber \\
 &  & +\frac{1}{4}\partial_{i}\left(g^{ji}\partial_{N}\mathcal{H}_{M\mathbf{2}\left(j\right)}\right)+\frac{1}{4}\partial_{i}\left(g^{ji}\partial_{M}\mathcal{H}_{N\mathbf{2}\left(j\right)}\right)\nonumber \\
 &  & +\frac{1}{4}\tilde{\partial}^{i}\left(\mathcal{H}_{\quad M}^{\mathbf{1}\left(j\right)}\tilde{\partial}^{j}\mathcal{H}_{\quad N}^{\mathbf{1}\left(i\right)}\right)+\frac{1}{4}\tilde{\partial}^{i}\left(\mathcal{H}_{\quad N}^{\mathbf{1}\left(j\right)}\tilde{\partial}^{j}\mathcal{H}_{\quad M}^{\mathbf{1}\left(i\right)}\right)\nonumber \\
 &  & +\frac{1}{4}\tilde{\partial}^{i}\left(\mathcal{H}_{\quad M}^{\mathbf{2}\left(j\right)}\partial_{j}\mathcal{H}_{\quad N}^{\mathbf{1}\left(i\right)}\right)+\frac{1}{4}\tilde{\partial}^{i}\left(\mathcal{H}_{\quad N}^{\mathbf{2}\left(j\right)}\partial_{j}\mathcal{H}_{\quad M}^{\mathbf{1}\left(i\right)}\right)\nonumber \\
 &  & +\frac{1}{4}\partial_{i}\left(\mathcal{H}_{\quad M}^{\mathbf{1}\left(j\right)}\tilde{\partial}^{j}\mathcal{H}_{\quad N}^{\mathbf{2}\left(i\right)}\right)+\frac{1}{4}\partial_{i}\left(\mathcal{H}_{\quad N}^{\mathbf{1}\left(j\right)}\tilde{\partial}^{j}\mathcal{H}_{\quad M}^{\mathbf{2}\left(i\right)}\right)\nonumber \\
 &  & +\frac{1}{4}\partial_{i}\left(\mathcal{H}_{\quad M}^{\mathbf{2}\left(j\right)}\partial_{j}\mathcal{H}_{\quad N}^{\mathbf{2}\left(i\right)}\right)+\frac{1}{4}\partial_{i}\left(\mathcal{H}_{\quad N}^{\mathbf{2}\left(j\right)}\partial_{j}\mathcal{H}_{\quad M}^{\mathbf{2}\left(i\right)}\right)\nonumber \\
 &  & -\frac{1}{8}S_{\quad M}^{\mathbf{1}\left(i\right)}\tilde{\partial}^{i}\tilde{g}_{kl}\tilde{\partial}^{j}\tilde{g}^{kl}S_{\quad N}^{\mathbf{1}\left(j\right)}-\frac{1}{8}S_{\quad M}^{\mathbf{1}\left(i\right)}\tilde{\partial}^{i}g^{kl}\tilde{\partial}^{j}g_{kl}S_{\quad N}^{\mathbf{1}\left(j\right)}\nonumber \\
 &  & -\frac{1}{8}S_{\quad M}^{\mathbf{1}\left(i\right)}\tilde{\partial}^{i}\tilde{g}_{kl}\partial_{j}\tilde{g}^{kl}S_{\quad N}^{\mathbf{2}\left(j\right)}-\frac{1}{8}S_{\quad M}^{\mathbf{1}\left(i\right)}\tilde{\partial}^{i}g^{kl}\partial_{j}g_{kl}S_{\quad N}^{\mathbf{2}\left(j\right)}\nonumber \\
 &  & -\frac{1}{8}S_{\quad M}^{\mathbf{2}\left(i\right)}\partial_{i}\tilde{g}_{kl}\tilde{\partial}^{j}\tilde{g}^{kl}S_{\quad N}^{\mathbf{1}\left(j\right)}-\frac{1}{8}S_{\quad M}^{\mathbf{2}\left(i\right)}\partial_{i}g^{kl}\tilde{\partial}^{j}g_{kl}S_{\quad N}^{\mathbf{1}\left(j\right)}\nonumber \\
 &  & -\frac{1}{8}S_{\quad M}^{\mathbf{2}\left(i\right)}\partial_{i}\tilde{g}_{kl}\partial_{j}\tilde{g}^{kl}S_{\quad N}^{\mathbf{2}\left(j\right)}-\frac{1}{8}S_{\quad M}^{\mathbf{2}\left(i\right)}\partial_{i}g^{kl}\partial_{j}g_{kl}S_{\quad N}^{\mathbf{2}\left(j\right)}\nonumber \\
 &  & +\frac{1}{4}S_{\quad M}^{\mathbf{1}\left(i\right)}\tilde{\partial}^{l}\left(\tilde{g}_{lk}\tilde{\partial}^{k}\tilde{g}^{ij}\right)S_{\quad N}^{\mathbf{1}\left(j\right)}+\frac{1}{4}S_{\quad M}^{\mathbf{1}\left(i\right)}\partial_{l}\left(g^{lk}\partial_{k}\tilde{g}^{ij}\right)S_{\quad N}^{\mathbf{1}\left(j\right)}\nonumber \\
 &  & +\frac{1}{4}S_{\quad M}^{\mathbf{2}\left(i\right)}\tilde{\partial}^{l}\left(\tilde{g}_{lk}\tilde{\partial}^{k}g_{ij}\right)S_{\quad N}^{\mathbf{2}\left(j\right)}+\frac{1}{4}S_{\quad M}^{\mathbf{2}\left(i\right)}\partial_{l}\left(g^{lk}\partial_{k}g_{ij}\right)S_{\quad N}^{\mathbf{2}\left(j\right)}\nonumber \\
 &  & +\frac{1}{4}S_{\quad M}^{\mathbf{1}\left(i\right)}\tilde{\partial}^{j}\tilde{g}_{kl}\tilde{\partial}^{l}\tilde{g}^{ik}S_{\quad N}^{\mathbf{1}\left(j\right)}+\frac{1}{4}S_{\quad M}^{\mathbf{1}\left(i\right)}\tilde{\partial}^{i}\tilde{g}_{kl}\tilde{\partial}^{l}\tilde{g}^{jk}S_{\quad N}^{\mathbf{1}\left(j\right)}\nonumber \\
 &  & +\frac{1}{4}S_{\quad M}^{\mathbf{1}\left(i\right)}\partial_{j}\tilde{g}_{kl}\tilde{\partial}^{l}\tilde{g}^{ik}S_{\quad N}^{\mathbf{2}\left(j\right)}+\frac{1}{4}S_{\quad M}^{\mathbf{1}\left(i\right)}\tilde{\partial}^{i}g^{kl}\partial_{l}g_{jk}S_{\quad N}^{\mathbf{2}\left(j\right)}\nonumber \\
 &  & +\frac{1}{4}S_{\quad M}^{\mathbf{2}\left(i\right)}\partial_{i}\tilde{g}_{kl}\tilde{\partial}^{l}\tilde{g}^{jk}S_{\quad N}^{\mathbf{1}\left(j\right)}+\frac{1}{4}S_{\quad M}^{\mathbf{2}\left(i\right)}\tilde{\partial}^{j}g^{kl}\partial_{l}g_{ik}S_{\quad N}^{\mathbf{1}\left(j\right)}\nonumber \\
 &  & +\frac{1}{4}S_{\quad M}^{\mathbf{2}\left(i\right)}\partial_{j}g^{kl}\partial_{l}g_{ik}S_{\quad N}^{\mathbf{2}\left(j\right)}+\frac{1}{4}S_{\quad M}^{\mathbf{2}\left(i\right)}\partial_{i}g^{kl}\partial_{l}g_{jk}S_{\quad N}^{\mathbf{2}\left(j\right)}\nonumber \\
 &  & -\frac{1}{4}S_{\quad M}^{\mathbf{1}\left(i\right)}\tilde{\partial}^{l}\left(\tilde{g}_{kl}\tilde{\partial}^{j}\tilde{g}^{ik}\right)S_{\quad N}^{\mathbf{1}\left(j\right)}-\frac{1}{4}S_{\quad M}^{\mathbf{1}\left(i\right)}\tilde{\partial}^{l}\left(\tilde{g}_{kl}\tilde{\partial}^{i}\tilde{g}^{jk}\right)S_{\quad N}^{\mathbf{1}\left(j\right)}\nonumber \\
 &  & -\frac{1}{4}S_{\quad M}^{\mathbf{1}\left(i\right)}\tilde{\partial}^{l}\left(\tilde{g}_{kl}\partial_{j}\tilde{g}^{ik}\right)S_{\quad N}^{\mathbf{2}\left(j\right)}-\frac{1}{4}S_{\quad M}^{\mathbf{1}\left(i\right)}\partial_{l}\left(g^{kl}\tilde{\partial}^{i}g_{jk}\right)S_{\quad N}^{\mathbf{2}\left(j\right)}\nonumber \\
 &  & -\frac{1}{4}S_{\quad M}^{\mathbf{2}\left(i\right)}\tilde{\partial}^{l}\left(\tilde{g}_{kl}\partial_{i}\tilde{g}^{jk}\right)S_{\quad N}^{\mathbf{1}\left(j\right)}-\frac{1}{4}S_{\quad M}^{\mathbf{2}\left(i\right)}\partial_{l}\left(g^{kl}\tilde{\partial}^{j}g_{ik}\right)S_{\quad N}^{\mathbf{1}\left(j\right)}\nonumber \\
 &  & -\frac{1}{4}S_{\quad M}^{\mathbf{2}\left(i\right)}\partial_{l}\left(g^{kl}\partial_{j}g_{ik}\right)S_{\quad N}^{\mathbf{2}\left(j\right)}-\frac{1}{4}S_{\quad M}^{\mathbf{2}\left(i\right)}\partial_{l}\left(g^{kl}\partial_{i}g_{jk}\right)S_{\quad N}^{\mathbf{2}\left(j\right)}\nonumber \\
 &  & -\frac{1}{4}S_{\quad M}^{\mathbf{1}\left(i\right)}\partial_{l}\left(\tilde{g}^{ki}\partial_{k}\tilde{g}^{lj}\right)S_{\quad N}^{\mathbf{1}\left(j\right)}-\frac{1}{4}S_{\quad M}^{\mathbf{1}\left(i\right)}\partial_{l}\left(\tilde{g}^{kj}\partial_{k}\tilde{g}^{li}\right)S_{\quad N}^{\mathbf{1}\left(j\right)}\nonumber \\
 &  & -\frac{1}{4}S_{\quad M}^{\mathbf{1}\left(i\right)}\tilde{\partial}^{l}\left(\tilde{g}^{ki}\partial_{k}g_{lj}\right)S_{\quad N}^{\mathbf{2}\left(j\right)}-\frac{1}{4}S_{\quad M}^{\mathbf{2}\left(i\right)}\partial_{l}\left(g_{ki}\tilde{\partial}^{k}\tilde{g}^{lj}\right)S_{\quad N}^{\mathbf{1}\left(j\right)}\nonumber \\
 &  & -\frac{1}{4}S_{\quad M}^{\mathbf{2}\left(i\right)}\tilde{\partial}^{l}\left(\tilde{g}^{kj}\partial_{k}g_{li}\right)S_{\quad N}^{\mathbf{1}\left(j\right)}-\frac{1}{4}S_{\quad M}^{\mathbf{1}\left(i\right)}\partial_{l}\left(g_{kj}\tilde{\partial}^{k}\tilde{g}^{li}\right)S_{\quad N}^{\mathbf{2}\left(j\right)}\nonumber \\
 &  & -\frac{1}{4}S_{\quad M}^{\mathbf{2}\left(i\right)}\tilde{\partial}^{l}\left(g_{ki}\tilde{\partial}^{k}g_{lj}\right)S_{\quad N}^{\mathbf{2}\left(j\right)}-\frac{1}{4}S_{\quad M}^{\mathbf{2}\left(i\right)}\tilde{\partial}^{l}\left(g_{kj}\tilde{\partial}^{k}g_{li}\right)S_{\quad N}^{\mathbf{2}\left(j\right)}.
\end{eqnarray}

\noindent The components are

\noindent
\begin{eqnarray}
\star\mathcal{R}_{\mathbf{1}\left(p\right)\mathbf{1}\left(q\right)} & = & \frac{1}{8}\tilde{\partial}^{p}\tilde{g}_{ij}\tilde{\partial}^{q}\tilde{g}^{ij}+\frac{1}{8}\tilde{\partial}^{p}\tilde{g}^{ij}\tilde{\partial}^{q}\tilde{g}_{ij}-\frac{1}{4}\tilde{\partial}^{i}\left(\tilde{g}_{ij}\tilde{\partial}^{j}\tilde{g}^{pq}\right)-\frac{1}{4}\partial_{i}\left(g^{ij}\partial_{j}g^{pq}\right)\nonumber \\
 &  & -\frac{1}{4}\tilde{\partial}^{q}\tilde{g}_{ij}\tilde{\partial}^{j}\tilde{g}^{pi}-\frac{1}{4}\tilde{\partial}^{p}\tilde{g}_{ij}\tilde{\partial}^{j}\tilde{g}^{qi}+\frac{1}{4}\tilde{\partial}^{i}\left(\tilde{g}_{ji}\tilde{\partial}^{q}\tilde{g}^{pj}\right)+\frac{1}{4}\tilde{\partial}^{i}\left(\tilde{g}_{ji}\tilde{\partial}^{p}\tilde{g}^{qj}\right)\nonumber \\
 &  & +\frac{1}{4}\partial_{i}\left(g^{jp}\partial_{j}g^{iq}\right)+\frac{1}{4}\partial_{i}\left(g^{jq}\partial_{j}g^{ip}\right)-\frac{1}{8}g^{ip}\partial_{i}g_{kl}\partial_{j}g^{kl}g^{jq}\nonumber \\
 &  & -\frac{1}{8}g^{ip}\partial_{i}g^{kl}\partial_{j}g_{kl}g^{jq}+\frac{1}{4}\tilde{g}^{ip}\tilde{\partial}^{l}\left(\tilde{g}_{lk}\tilde{\partial}^{k}\tilde{g}_{ij}\right)\tilde{g}^{jq}+\frac{1}{4}g^{ip}\partial_{l}\left(g^{lk}\partial_{k}g_{ij}\right)g^{jq}\nonumber \\
 &  & +\frac{1}{4}g^{ip}\partial_{j}g^{kl}\partial_{l}g_{ik}g^{jq}+\frac{1}{4}g^{ip}\partial_{i}g^{kl}\partial_{l}g_{jk}g^{jq}-\frac{1}{4}g^{ip}\partial_{l}\left(g^{kl}\partial_{j}g_{ik}\right)g^{jq}\nonumber \\
 &  & -\frac{1}{4}g^{ip}\partial_{l}\left(g^{kl}\partial_{i}g_{jk}\right)g^{jq}-\frac{1}{4}\tilde{g}^{ip}\tilde{\partial}^{l}\left(\tilde{g}_{ki}\tilde{\partial}^{k}\tilde{g}_{lj}\right)\tilde{g}^{jq}-\frac{1}{4}\tilde{g}^{ip}\tilde{\partial}^{l}\left(\tilde{g}_{kj}\tilde{\partial}^{k}\tilde{g}_{li}\right)\tilde{g}^{jq},\nonumber \\
\nonumber \\
\star\mathcal{R}_{\mathbf{2}\left(p\right)\mathbf{2}\left(q\right)} & = & \frac{1}{8}\partial_{p}g_{ij}\partial_{q}g^{ij}+\frac{1}{8}\partial_{p}g^{ij}\partial_{q}g_{ij}-\frac{1}{4}\tilde{\partial}^{i}\left(\tilde{g}_{ij}\tilde{\partial}^{j}\tilde{g}_{pq}\right)-\frac{1}{4}\partial_{i}\left(g^{ij}\partial_{j}g_{pq}\right)\nonumber \\
 &  & -\frac{1}{4}\partial_{q}g^{ij}\partial_{j}g_{pi}-\frac{1}{4}\partial_{p}g^{ij}\partial_{j}g_{qi}+\frac{1}{4}\partial_{i}\left(g^{ji}\partial_{q}g_{pj}\right)+\frac{1}{4}\partial_{i}\left(g^{ji}\partial_{p}g_{qj}\right)\nonumber \\
 &  & +\frac{1}{4}\tilde{\partial}^{i}\left(\tilde{g}_{jp}\tilde{\partial}^{j}\tilde{g}_{iq}\right)+\frac{1}{4}\tilde{\partial}^{i}\left(\tilde{g}_{jq}\tilde{\partial}^{j}\tilde{g}_{ip}\right)-\frac{1}{8}\tilde{g}_{ip}\tilde{\partial}^{i}\tilde{g}_{kl}\tilde{\partial}^{j}\tilde{g}^{kl}\tilde{g}_{jq}\nonumber \\
 &  & -\frac{1}{8}\tilde{g}_{ip}\tilde{\partial}^{i}\tilde{g}^{kl}\tilde{\partial}^{j}\tilde{g}_{kl}\tilde{g}_{jq}+\frac{1}{4}\tilde{g}_{ip}\tilde{\partial}^{l}\left(\tilde{g}_{lk}\tilde{\partial}^{k}\tilde{g}^{ij}\right)\tilde{g}_{jq}+\frac{1}{4}g_{ip}\partial_{l}\left(g^{lk}\partial_{k}g^{ij}\right)g_{jq}\nonumber \\
 &  & +\frac{1}{4}\tilde{g}_{ip}\tilde{\partial}^{j}\tilde{g}_{kl}\tilde{\partial}^{l}\tilde{g}^{ik}\tilde{g}_{jq}+\frac{1}{4}\tilde{g}_{ip}\tilde{\partial}^{i}\tilde{g}_{kl}\tilde{\partial}^{l}\tilde{g}^{jk}\tilde{g}_{jq}-\frac{1}{4}\tilde{g}_{ip}\tilde{\partial}^{l}\left(\tilde{g}_{kl}\tilde{\partial}^{j}\tilde{g}^{ik}\right)\tilde{g}_{jq}\nonumber \\
 &  & -\frac{1}{4}\tilde{g}_{ip}\tilde{\partial}^{l}\left(\tilde{g}_{kl}\tilde{\partial}^{i}\tilde{g}^{jk}\right)\tilde{g}_{jq}-\frac{1}{4}g_{ip}\partial_{l}\left(g^{ki}\partial_{k}g^{lj}\right)g_{jq}-\frac{1}{4}g_{ip}\partial_{l}\left(g^{kj}\partial_{k}g^{li}\right)g_{jq},\nonumber \\
\nonumber \\
\star\mathcal{R}_{\mathbf{1}\left(p\right)\mathbf{2}\left(q\right)} & = & \frac{1}{8}\tilde{\partial}^{p}\tilde{g}_{ij}\partial_{q}g^{ij}+\frac{1}{8}\tilde{\partial}^{p}\tilde{g}^{ij}\partial_{q}g_{ij}-\frac{1}{4}\partial_{q}g_{ij}\tilde{\partial}^{j}\tilde{g}^{pi}-\frac{1}{4}\tilde{\partial}^{p}\tilde{g}^{ij}\partial_{j}g_{qi}\nonumber \\
 &  & +\frac{1}{4}\tilde{\partial}^{i}\left(\tilde{g}_{ji}\partial_{q}\tilde{g}^{pj}\right)+\frac{1}{4}\partial_{i}\left(g^{ji}\tilde{\partial}^{p}g_{qj}\right)+\frac{1}{4}\tilde{\partial}^{i}\left(\tilde{g}^{jp}\partial_{j}g_{iq}\right)+\frac{1}{4}\partial_{i}\left(g_{jq}\tilde{\partial}^{j}\tilde{g}^{ip}\right)\nonumber \\
 &  & -\frac{1}{8}g^{ip}\partial_{i}g_{kl}\tilde{\partial}^{j}\tilde{g}^{kl}g_{jq}-\frac{1}{8}g^{ip}\partial_{i}g^{kl}\tilde{\partial}^{j}\tilde{g}_{kl}g_{jq}+\frac{1}{4}g^{ip}\partial_{i}g_{kl}\tilde{\partial}^{l}\tilde{g}^{jk}g_{jq}\nonumber \\
 &  & +\frac{1}{4}g^{ip}\tilde{\partial}^{j}\tilde{g}^{kl}\partial_{l}g_{ik}g_{jq}-\frac{1}{4}g^{ip}\tilde{\partial}^{l}\left(\tilde{g}_{kl}\partial_{i}g^{jk}\right)g_{jq}-\frac{1}{4}g^{ip}\partial_{l}\left(g^{kl}\tilde{\partial}^{j}\tilde{g}_{ik}\right)g_{jq}\nonumber \\
 &  & -\frac{1}{4}g^{ip}\partial_{l}\left(g_{ki}\tilde{\partial}^{k}\tilde{g}^{lj}\right)g_{jq}-\frac{1}{4}g^{ip}\tilde{\partial}^{l}\left(g^{kj}\partial_{k}g_{li}\right)g_{jq},\nonumber \\
\nonumber \\
\star\mathcal{R}_{\mathbf{2}\left(p\right)\mathbf{1}\left(q\right)} & = & \frac{1}{8}\partial_{p}g_{ij}\tilde{\partial}^{q}\tilde{g}^{ij}+\frac{1}{8}\partial_{p}g^{ij}\tilde{\partial}^{q}\tilde{g}_{ij}-\frac{1}{4}\partial_{p}g_{ij}\tilde{\partial}^{j}\tilde{g}^{qi}-\frac{1}{4}\tilde{\partial}^{q}\tilde{g}^{ij}\partial_{j}g_{pi}\nonumber \\
 &  & +\frac{1}{4}\tilde{\partial}^{i}\left(g_{ji}\partial_{p}g^{qj}\right)+\frac{1}{4}\partial_{i}\left(g^{ji}\tilde{\partial}^{q}\tilde{g}_{pj}\right)+\frac{1}{4}\tilde{\partial}^{i}\left(g^{jq}\partial_{j}g_{ip}\right)+\frac{1}{4}\partial_{i}\left(g_{jp}\tilde{\partial}^{j}\tilde{g}^{iq}\right)\nonumber \\
 &  & -\frac{1}{8}g_{ip}\tilde{\partial}^{i}\tilde{g}_{kl}\partial_{j}g^{kl}g^{jq}-\frac{1}{8}g_{ip}\tilde{\partial}^{i}\tilde{g}^{kl}\partial_{j}g_{kl}g^{jq}+\frac{1}{4}g_{ip}\partial_{j}g_{kl}\tilde{\partial}^{l}\tilde{g}^{ik}g^{jq}\nonumber \\
 &  & +\frac{1}{4}g_{ip}\tilde{\partial}^{i}\tilde{g}^{kl}\partial_{l}g_{jk}g^{jq}-\frac{1}{4}g_{ip}\tilde{\partial}^{l}\left(g_{kl}\partial_{j}g^{ik}\right)g^{jq}-\frac{1}{4}g_{ip}\partial_{l}\left(g^{kl}\tilde{\partial}^{i}\tilde{g}_{jk}\right)g^{jq}\nonumber \\
 &  & -\frac{1}{4}g_{ip}\tilde{\partial}^{l}\left(g^{ki}\partial_{k}g_{lj}\right)g^{jq}-\frac{1}{4}g_{ip}\partial_{l}\left(g_{kj}\tilde{\partial}^{k}\tilde{g}^{li}\right)g^{jq}.
\end{eqnarray}

\noindent It is straightforward to verify the symmetry

\noindent
\begin{eqnarray}
\star\mathcal{R}_{\mathbf{1}\left(p\right)\mathbf{1}\left(q\right)} & \underleftrightarrow{g^{\bullet\bullet}\leftrightarrow g_{\bullet\bullet},\quad\tilde{\partial}^{\bullet}\leftrightarrow\partial_{\bullet}} & \star\mathcal{R}_{\mathbf{2}\left(p\right)\mathbf{2}\left(q\right)},\nonumber \\
\star\mathcal{R}_{\mathbf{1}\left(p\right)\mathbf{2}\left(q\right)}
& \underleftrightarrow{g^{\bullet\bullet}\leftrightarrow
g_{\bullet\bullet},\quad\tilde{\partial}^{\bullet}\leftrightarrow\partial_{\bullet}}
&
\star\mathcal{R}_{\mathbf{2}\left(p\right)\mathbf{1}\left(q\right)}.
\end{eqnarray}

\noindent With our metric ansatz (\ref{MetricAnsatz}) and
calculation rules in section (4.1), we obtain

\noindent
\begin{eqnarray}
\star\mathcal{R}_{\mathbf{2}\left(t\right)\mathbf{2}\left(t\right)} & = & -\left(D-1\right)\frac{\dot{a}^{2}}{a^{2}}+\left(D-1\right)\frac{\dot{\tilde{a}}^{2}}{\tilde{a}^{2}},\nonumber \\
\star\mathcal{R}_{\mathbf{2}\left(i\right)\mathbf{2}\left(i\right)} & = & \frac{1}{\tilde{a}^{4}}\left(\dot{\tilde{a}}^{2}-\tilde{a}\ddot{\tilde{a}}\right)-\left(\dot{a}^{2}-a\ddot{a}\right),\nonumber \\
\star\mathcal{R}_{\mathbf{1}\left(t\right)\mathbf{2}\left(t\right)}
& = & 0.
\end{eqnarray}

\section*{\noindent \underline{Calculation of
$*\mathcal{R}_{MN}$}}

\noindent
\begin{eqnarray}
*\mathcal{R}_{MN} & = & *\mathcal{K}_{MN}-S_{\quad M}^{P}*\mathcal{K}_{PQ}S_{\quad N}^{Q}\nonumber \\
 & = & \frac{1}{2}\left(-\frac{1}{4}\tilde{g}_{ab}\tilde{\partial}^{i}\tilde{g}^{ab}+\tilde{\partial}^{i}\tilde{\phi}\right)\left(\tilde{g}_{ij}\tilde{\partial}^{j}\mathcal{H}_{MN}\right)\nonumber \\
 &  & +\frac{1}{2}\left(-\frac{1}{4}g^{ab}\partial_{i}g_{ab}+\partial_{i}\phi\right)\left(g^{ij}\partial_{j}\mathcal{H}_{MN}\right)+2\partial_{M}\partial_{N}d\nonumber \\
 &  & -\left(-\frac{1}{4}\tilde{g}_{ab}\tilde{\partial}^{i}\tilde{g}^{ab}+\tilde{\partial}^{i}\tilde{\phi}\right)\left(\mathcal{H}_{\quad\left(M\right.}^{\mathbf{1}\left(j\right)}\tilde{\partial}^{j}\mathcal{H}_{\quad\left.N\right)}^{\mathbf{1}\left(i\right)}\right)\nonumber \\
 &  & -\left(-\frac{1}{4}\tilde{g}_{ab}\tilde{\partial}^{i}\tilde{g}^{ab}+\tilde{\partial}^{i}\tilde{\phi}\right)\left(\mathcal{H}_{\quad\left(M\right.}^{\mathbf{2}\left(j\right)}\partial_{j}\mathcal{H}_{\quad\left.N\right)}^{\mathbf{1}\left(i\right)}\right)\nonumber \\
 &  & -\left(-\frac{1}{4}g^{ab}\partial_{i}g_{ab}+\partial_{i}\phi\right)\left(\mathcal{H}_{\quad\left(M\right.}^{\mathbf{1}\left(j\right)}\tilde{\partial}^{j}\mathcal{H}_{\quad\left.N\right)}^{\mathbf{2}\left(i\right)}\right)\nonumber \\
 &  & -\left(-\frac{1}{4}g^{ab}\partial_{i}g_{ab}+\partial_{i}\phi\right)\left(\mathcal{H}_{\quad\left(M\right.}^{\mathbf{2}\left(j\right)}\partial_{j}\mathcal{H}_{\quad\left.N\right)}^{\mathbf{2}\left(i\right)}\right)\nonumber \\
 &  & -\left(-\frac{1}{4}\tilde{g}_{ab}\tilde{\partial}^{i}\tilde{g}^{ab}+\tilde{\partial}^{i}\tilde{\phi}\right)\left(\tilde{g}_{ji}\partial_{\left(N\right.}\mathcal{H}_{\left.M\right)\mathbf{1}\left(j\right)}\right)\nonumber \\
 &  & -\left(-\frac{1}{4}g^{ab}\partial_{i}g_{ab}+\partial_{i}\phi\right)\left(g^{ji}\partial_{\left(N\right.}\mathcal{H}_{\left.M\right)\mathbf{2}\left(j\right)}\right)\nonumber \\
 &  & -\frac{1}{2}S_{\quad M}^{\mathbf{1}\left(i\right)}\left(-\frac{1}{4}\tilde{g}_{ab}\tilde{\partial}^{l}\tilde{g}^{ab}+\tilde{\partial}^{l}\tilde{\phi}\right)\left(\tilde{g}_{lk}\tilde{\partial}^{k}\tilde{g}^{ij}\right)S_{\quad N}^{\mathbf{1}\left(j\right)}\nonumber \\
 &  & -\frac{1}{2}S_{\quad M}^{\mathbf{1}\left(i\right)}\left(-\frac{1}{4}g^{ab}\partial_{l}g_{ab}+\partial_{l}\phi\right)\left(g^{lk}\partial_{k}\tilde{g}^{ij}\right)S_{\quad N}^{\mathbf{1}\left(j\right)}\nonumber \\
 &  & -\frac{1}{2}S_{\quad M}^{\mathbf{2}\left(i\right)}\left(-\frac{1}{4}\tilde{g}_{ab}\tilde{\partial}^{l}\tilde{g}^{ab}+\tilde{\partial}^{l}\tilde{\phi}\right)\left(\tilde{g}_{lk}\tilde{\partial}^{k}g_{ij}\right)S_{\quad N}^{\mathbf{2}\left(j\right)}\nonumber \\
 &  & -\frac{1}{2}S_{\quad M}^{\mathbf{2}\left(i\right)}\left(-\frac{1}{4}g^{ab}\partial_{l}g_{ab}+\partial_{l}\phi\right)\left(g^{lk}\partial_{k}g_{ij}\right)S_{\quad N}^{\mathbf{2}\left(j\right)}\nonumber \\
 &  & -2S_{\quad M}^{\mathbf{1}\left(i\right)}\partial_{\mathbf{1}\left(i\right)}\partial_{\mathbf{1}\left(j\right)}dS_{\quad N}^{\mathbf{1}\left(j\right)}-2S_{\quad M}^{\mathbf{1}\left(i\right)}\partial_{\mathbf{1}\left(i\right)}\partial_{\mathbf{2}\left(j\right)}dS_{\quad N}^{\mathbf{2}\left(j\right)}\nonumber \\
 &  & -2S_{\quad M}^{\mathbf{2}\left(i\right)}\partial_{\mathbf{2}\left(i\right)}\partial_{\mathbf{1}\left(j\right)}dS_{\quad N}^{\mathbf{1}\left(j\right)}-2S_{\quad M}^{\mathbf{2}\left(i\right)}\partial_{\mathbf{2}\left(i\right)}\partial_{\mathbf{2}\left(j\right)}dS_{\quad N}^{\mathbf{2}\left(j\right)}\nonumber \\
 &  & +\frac{1}{2}S_{\quad M}^{\mathbf{1}\left(i\right)}\left(-\frac{1}{4}\tilde{g}_{ab}\tilde{\partial}^{l}\tilde{g}^{ab}+\tilde{\partial}^{l}\tilde{\phi}\right)\left(\tilde{g}_{kl}\tilde{\partial}^{j}\tilde{g}^{ik}\right)S_{\quad N}^{\mathbf{1}\left(j\right)}\nonumber \\
 &  & +\frac{1}{2}S_{\quad M}^{\mathbf{1}\left(i\right)}\left(-\frac{1}{4}\tilde{g}_{ab}\tilde{\partial}^{l}\tilde{g}^{ab}+\tilde{\partial}^{l}\tilde{\phi}\right)\left(\tilde{g}_{kl}\tilde{\partial}^{i}\tilde{g}^{jk}\right)S_{\quad N}^{\mathbf{1}\left(j\right)}\nonumber \\
 &  & +\frac{1}{2}S_{\quad M}^{\mathbf{1}\left(i\right)}\left(-\frac{1}{4}\tilde{g}_{ab}\tilde{\partial}^{l}\tilde{g}^{ab}+\tilde{\partial}^{l}\tilde{\phi}\right)\left(\tilde{g}_{kl}\partial_{j}\tilde{g}^{ik}\right)S_{\quad N}^{\mathbf{2}\left(j\right)}\nonumber \\
 &  & +\frac{1}{2}S_{\quad M}^{\mathbf{1}\left(i\right)}\left(-\frac{1}{4}g^{ab}\partial_{l}g_{ab}+\partial_{l}\phi\right)\left(g^{kl}\tilde{\partial}^{i}g_{jk}\right)S_{\quad N}^{\mathbf{2}\left(j\right)}\nonumber \\
 &  & +\frac{1}{2}S_{\quad M}^{\mathbf{2}\left(i\right)}\left(-\frac{1}{4}\tilde{g}_{ab}\tilde{\partial}^{l}\tilde{g}^{ab}+\tilde{\partial}^{l}\tilde{\phi}\right)\left(\tilde{g}_{kl}\partial_{i}\tilde{g}^{jk}\right)S_{\quad N}^{\mathbf{1}\left(j\right)}\nonumber \\
 &  & +\frac{1}{2}S_{\quad M}^{\mathbf{2}\left(i\right)}\left(-\frac{1}{4}g^{ab}\partial_{l}g_{ab}+\partial_{l}\phi\right)\left(g^{kl}\tilde{\partial}^{j}g_{ik}\right)S_{\quad N}^{\mathbf{1}\left(j\right)}\nonumber \\
 &  & +\frac{1}{2}S_{\quad M}^{\mathbf{2}\left(i\right)}\left(-\frac{1}{4}g^{ab}\partial_{l}g_{ab}+\partial_{l}\phi\right)\left(g^{kl}\partial_{j}g_{ik}\right)S_{\quad N}^{\mathbf{2}\left(j\right)}\nonumber \\
 &  & +\frac{1}{2}S_{\quad M}^{\mathbf{2}\left(i\right)}\left(-\frac{1}{4}g^{ab}\partial_{l}g_{ab}+\partial_{l}\phi\right)\left(g^{kl}\partial_{i}g_{jk}\right)S_{\quad N}^{\mathbf{2}\left(j\right)}\nonumber \\
 &  & +\frac{1}{2}S_{\quad M}^{\mathbf{1}\left(i\right)}\left(-\frac{1}{4}g^{ab}\partial_{l}g_{ab}+\partial_{l}\phi\right)\left(\tilde{g}^{ki}\partial_{k}\tilde{g}^{lj}\right)S_{\quad N}^{\mathbf{1}\left(j\right)}\nonumber \\
 &  & +\frac{1}{2}S_{\quad M}^{\mathbf{1}\left(i\right)}\left(-\frac{1}{4}g^{ab}\partial_{l}g_{ab}+\partial_{l}\phi\right)\left(\tilde{g}^{kj}\partial_{k}\tilde{g}^{li}\right)S_{\quad N}^{\mathbf{1}\left(j\right)}\nonumber \\
 &  & +\frac{1}{2}S_{\quad M}^{\mathbf{1}\left(i\right)}\left(-\frac{1}{4}\tilde{g}_{ab}\tilde{\partial}^{l}\tilde{g}^{ab}+\tilde{\partial}^{l}\tilde{\phi}\right)\left(\tilde{g}^{ki}\partial_{k}g_{lj}\right)S_{\quad N}^{\mathbf{2}\left(j\right)}\nonumber \\
 &  & +\frac{1}{2}S_{\quad M}^{\mathbf{1}\left(i\right)}\left(-\frac{1}{4}g^{ab}\partial_{l}g_{ab}+\partial_{l}\phi\right)\left(g_{kj}\tilde{\partial}^{k}\tilde{g}^{li}\right)S_{\quad N}^{\mathbf{2}\left(j\right)}\nonumber \\
 &  & +\frac{1}{2}S_{\quad M}^{\mathbf{2}\left(i\right)}\left(-\frac{1}{4}\tilde{g}_{ab}\tilde{\partial}^{l}\tilde{g}^{ab}+\tilde{\partial}^{l}\tilde{\phi}\right)\left(\tilde{g}^{kj}\partial_{k}g_{li}\right)S_{\quad N}^{\mathbf{1}\left(j\right)}\nonumber \\
 &  & +\frac{1}{2}S_{\quad M}^{\mathbf{2}\left(i\right)}\left(-\frac{1}{4}g^{ab}\partial_{l}g_{ab}+\partial_{l}\phi\right)\left(g_{ki}\tilde{\partial}^{k}\tilde{g}^{lj}\right)S_{\quad N}^{\mathbf{1}\left(j\right)}\nonumber \\
 &  & +\frac{1}{2}S_{\quad M}^{\mathbf{2}\left(i\right)}\left(-\frac{1}{4}\tilde{g}_{ab}\tilde{\partial}^{l}\tilde{g}^{ab}+\tilde{\partial}^{l}\tilde{\phi}\right)\left(g_{ki}\tilde{\partial}^{k}g_{lj}\right)S_{\quad N}^{\mathbf{2}\left(j\right)}\nonumber \\
 &  & +\frac{1}{2}S_{\quad M}^{\mathbf{2}\left(i\right)}\left(-\frac{1}{4}\tilde{g}_{ab}\tilde{\partial}^{l}\tilde{g}^{ab}+\tilde{\partial}^{l}\tilde{\phi}\right)\left(g_{kj}\tilde{\partial}^{k}g_{li}\right)S_{\quad N}^{\mathbf{2}\left(j\right)}.
\end{eqnarray}

\noindent Thus the components are

\noindent
\begin{eqnarray}
*\mathcal{R}_{\mathbf{1}\left(p\right)\mathbf{1}\left(q\right)} & = & \frac{1}{2}\left(-\frac{1}{4}\tilde{g}_{ab}\tilde{\partial}^{i}\tilde{g}^{ab}+\tilde{\partial}^{i}\tilde{\phi}\right)\left(\tilde{g}_{ij}\tilde{\partial}^{j}\tilde{g}^{pq}\right)+\frac{1}{2}\left(-\frac{1}{4}g^{ab}\partial_{i}g_{ab}+\partial_{i}\phi\right)\left(g^{ij}\partial_{j}\tilde{g}^{pq}\right)\nonumber \\
 &  & -\frac{1}{2}\tilde{\partial}^{p}\tilde{g}_{aa}\tilde{\partial}^{q}\tilde{g}^{aa}-\frac{1}{2}\tilde{g}_{aa}\tilde{\partial}^{p}\tilde{\partial}^{q}\tilde{g}^{aa}+\tilde{\partial}^{p}\tilde{\partial}^{q}\tilde{\phi}\nonumber \\
 &  & -2g^{ip}\left(-\frac{1}{4}\partial_{i}g^{aa}\partial_{j}g_{aa}-\frac{1}{4}g^{aa}\partial_{i}\partial_{j}g_{aa}+\partial_{i}\partial_{j}\phi\right)g^{jq}\nonumber \\
 &  & -\frac{1}{2}\left(-\frac{1}{4}g^{ab}\partial_{i}g_{ab}+\partial_{i}\phi\right)\left(g^{jp}\partial_{j}g^{iq}\right)-\frac{1}{2}\left(-\frac{1}{4}g^{ab}\partial_{i}g_{ab}+\partial_{i}\phi\right)\left(g^{jq}\partial_{j}g^{ip}\right)\nonumber \\
 &  & -\frac{1}{2}\left(-\frac{1}{4}\tilde{g}_{ab}\tilde{\partial}^{i}\tilde{g}^{ab}+\tilde{\partial}^{i}\tilde{\phi}\right)\left(\tilde{g}_{ji}\tilde{\partial}^{q}\tilde{g}^{pj}\right)-\frac{1}{2}\left(-\frac{1}{4}\tilde{g}_{ab}\tilde{\partial}^{i}\tilde{g}^{ab}+\tilde{\partial}^{i}\tilde{\phi}\right)\left(\tilde{g}_{ji}\tilde{\partial}^{p}\tilde{g}^{qj}\right)\nonumber \\
 &  & -\frac{1}{2}g^{ip}\left(-\frac{1}{4}\tilde{g}_{ab}\tilde{\partial}^{l}\tilde{g}^{ab}+\tilde{\partial}^{l}\tilde{\phi}\right)\left(\tilde{g}_{lk}\tilde{\partial}^{k}g_{ij}\right)g^{jq}\nonumber \\
 &  & -\frac{1}{2}g^{ip}\left(-\frac{1}{4}g^{ab}\partial_{l}g_{ab}+\partial_{l}\phi\right)\left(g^{lk}\partial_{k}g_{ij}\right)g^{jq}\nonumber \\
 &  & +\frac{1}{2}g^{ip}\left(-\frac{1}{4}g^{ab}\partial_{l}g_{ab}+\partial_{l}\phi\right)\left(g^{kl}\partial_{j}g_{ik}\right)g^{jq}\nonumber \\
 &  & +\frac{1}{2}g^{ip}\left(-\frac{1}{4}g^{ab}\partial_{l}g_{ab}+\partial_{l}\phi\right)\left(g^{kl}\partial_{i}g_{jk}\right)g^{jq}\nonumber \\
 &  & +\frac{1}{2}g^{ip}\left(-\frac{1}{4}\tilde{g}_{ab}\tilde{\partial}^{l}\tilde{g}^{ab}+\tilde{\partial}^{l}\tilde{\phi}\right)\left(g_{ki}\tilde{\partial}^{k}g_{lj}\right)g^{jq}\nonumber \\
 &  & +\frac{1}{2}g^{ip}\left(-\frac{1}{4}\tilde{g}_{ab}\tilde{\partial}^{l}\tilde{g}^{ab}+\tilde{\partial}^{l}\tilde{\phi}\right)\left(g_{kj}\tilde{\partial}^{k}g_{li}\right)g^{jq},\nonumber \\
\nonumber \\
*\mathcal{R}_{\mathbf{2}\left(p\right)\mathbf{2}\left(q\right)} & = & \frac{1}{2}\left(-\frac{1}{4}\tilde{g}_{ab}\tilde{\partial}^{i}\tilde{g}^{ab}+\tilde{\partial}^{i}\tilde{\phi}\right)\left(\tilde{g}_{ij}\tilde{\partial}^{j}g_{pq}\right)+\frac{1}{2}\left(-\frac{1}{4}g^{ab}\partial_{i}g_{ab}+\partial_{i}\phi\right)\left(g^{ij}\partial_{j}g_{pq}\right)\nonumber \\
 &  & -\frac{1}{2}\partial_{p}g^{aa}\partial_{q}g_{aa}-\frac{1}{2}g^{aa}\partial_{p}\partial_{q}g_{aa}+2\partial_{p}\partial_{q}\phi\nonumber \\
 &  & -2g_{ip}\left(-\frac{1}{4}\tilde{\partial}^{i}\tilde{g}_{aa}\tilde{\partial}^{j}\tilde{g}^{aa}-\frac{1}{4}\tilde{g}_{aa}\tilde{\partial}^{i}\tilde{\partial}^{j}\tilde{g}^{aa}+\tilde{\partial}^{i}\tilde{\partial}^{j}\tilde{\phi}\right)g_{jq}\nonumber \\
 &  & -\frac{1}{2}\left(-\frac{1}{4}\tilde{g}_{ab}\tilde{\partial}^{i}\tilde{g}^{ab}+\tilde{\partial}^{i}\tilde{\phi}\right)\left(g_{jp}\tilde{\partial}^{j}g_{iq}\right)-\frac{1}{2}\left(-\frac{1}{4}\tilde{g}_{ab}\tilde{\partial}^{i}\tilde{g}^{ab}+\tilde{\partial}^{i}\tilde{\phi}\right)\left(g_{jq}\tilde{\partial}^{j}g_{ip}\right)\nonumber \\
 &  & -\frac{1}{2}\left(-\frac{1}{4}g^{ab}\partial_{i}g_{ab}+\partial_{i}\phi\right)\left(g^{ji}\partial_{q}g_{pj}\right)-\frac{1}{2}\left(-\frac{1}{4}g^{ab}\partial_{i}g_{ab}+\partial_{i}\phi\right)\left(g^{ji}\partial_{p}g_{qj}\right)\nonumber \\
 &  & -\frac{1}{2}g_{ip}\left(-\frac{1}{4}\tilde{g}_{ab}\tilde{\partial}^{l}\tilde{g}^{ab}+\tilde{\partial}^{l}\tilde{\phi}\right)\left(\tilde{g}_{lk}\tilde{\partial}^{k}\tilde{g}^{ij}\right)g_{jq}\nonumber \\
 &  & -\frac{1}{2}g_{ip}\left(-\frac{1}{4}g^{ab}\partial_{l}g_{ab}+\partial_{l}\phi\right)\left(g^{lk}\partial_{k}\tilde{g}^{ij}\right)g_{jq}\nonumber \\
 &  & +\frac{1}{2}g_{ip}\left(-\frac{1}{4}\tilde{g}_{ab}\tilde{\partial}^{l}\tilde{g}^{ab}+\tilde{\partial}^{l}\tilde{\phi}\right)\left(\tilde{g}_{kl}\tilde{\partial}^{j}\tilde{g}^{ik}\right)g_{jq}\nonumber \\
 &  & +\frac{1}{2}g_{ip}\left(-\frac{1}{4}\tilde{g}_{ab}\tilde{\partial}^{l}\tilde{g}^{ab}+\tilde{\partial}^{l}\tilde{\phi}\right)\left(\tilde{g}_{kl}\tilde{\partial}^{i}\tilde{g}^{jk}\right)g_{jq}\nonumber \\
 &  & +\frac{1}{2}g_{ip}\left(-\frac{1}{4}g^{ab}\partial_{l}g_{ab}+\partial_{l}\phi\right)\left(\tilde{g}^{ki}\partial_{k}\tilde{g}^{lj}\right)g_{jq}\nonumber \\
 &  & +\frac{1}{2}g_{ip}\left(-\frac{1}{4}g^{ab}\partial_{l}g_{ab}+\partial_{l}\phi\right)\left(\tilde{g}^{kj}\partial_{k}\tilde{g}^{li}\right)g_{jq},\nonumber \\
\nonumber \\
*\mathcal{R}_{\mathbf{1}\left(p\right)\mathbf{2}\left(q\right)} & = & +2\tilde{\partial}^{p}\partial_{q}d-2\tilde{g}^{ip}\partial_{i}\tilde{\partial}^{j}dg_{jq}\nonumber \\
 &  & -\frac{1}{2}\left(-\frac{1}{4}\tilde{g}_{ab}\tilde{\partial}^{i}\tilde{g}^{ab}+\tilde{\partial}^{i}\tilde{\phi}\right)\left(\tilde{g}^{jp}\partial_{j}g_{iq}\right)-\frac{1}{2}\left(-\frac{1}{4}g^{ab}\partial_{i}g_{ab}+\partial_{i}\phi\right)\left(g_{jq}\tilde{\partial}^{j}\tilde{g}^{ip}\right)\nonumber \\
 &  & -\frac{1}{2}\left(-\frac{1}{4}\tilde{g}_{ab}\tilde{\partial}^{i}\tilde{g}^{ab}+\tilde{\partial}^{i}\tilde{\phi}\right)\left(\tilde{g}_{ji}\partial_{q}\tilde{g}^{pj}\right)-\frac{1}{2}\left(-\frac{1}{4}g^{ab}\partial_{i}g_{ab}+\partial_{i}\phi\right)\left(g^{ji}\tilde{\partial}^{p}g_{qj}\right)\nonumber \\
 &  & +\frac{1}{2}\tilde{g}^{ip}\left(-\frac{1}{4}\tilde{g}_{ab}\tilde{\partial}^{l}\tilde{g}^{ab}+\tilde{\partial}^{l}\tilde{\phi}\right)\left(\tilde{g}_{kl}\partial_{i}\tilde{g}^{jk}\right)g_{jq}\nonumber \\
 &  & +\frac{1}{2}\tilde{g}^{ip}\left(-\frac{1}{4}g^{ab}\partial_{l}g_{ab}+\partial_{l}\phi\right)\left(g^{kl}\tilde{\partial}^{j}g_{ik}\right)g_{jq}\nonumber \\
 &  & +\frac{1}{2}\tilde{g}^{ip}\left(-\frac{1}{4}\tilde{g}_{ab}\tilde{\partial}^{l}\tilde{g}^{ab}+\tilde{\partial}^{l}\tilde{\phi}\right)\left(\tilde{g}^{kj}\partial_{k}g_{li}\right)g_{jq}\nonumber \\
 &  & +\frac{1}{2}\tilde{g}^{ip}\left(-\frac{1}{4}g^{ab}\partial_{l}g_{ab}+\partial_{l}\phi\right)\left(g_{ki}\tilde{\partial}^{k}\tilde{g}^{lj}\right)g_{jq},\nonumber \\
\nonumber \\
*\mathcal{R}_{\mathbf{2}\left(p\right)\mathbf{1}\left(q\right)} & = & +2\partial_{p}\tilde{\partial}^{q}d-2\tilde{g}_{ip}\tilde{\partial}^{i}\partial_{j}dg^{jq}\nonumber \\
 &  & -\frac{1}{2}\left(-\frac{1}{4}\tilde{g}_{ab}\tilde{\partial}^{i}\tilde{g}^{ab}+\tilde{\partial}^{i}\tilde{\phi}\right)\left(g^{jq}\partial_{j}\tilde{g}_{ip}\right)-\frac{1}{2}\left(-\frac{1}{4}g^{ab}\partial_{i}g_{ab}+\partial_{i}\phi\right)\left(\tilde{g}_{jp}\tilde{\partial}^{j}g^{iq}\right)\nonumber \\
 &  & -\frac{1}{2}\left(-\frac{1}{4}\tilde{g}_{ab}\tilde{\partial}^{i}\tilde{g}^{ab}+\tilde{\partial}^{i}\tilde{\phi}\right)\left(\tilde{g}_{ji}\partial_{p}\tilde{g}^{qj}\right)-\frac{1}{2}\left(-\frac{1}{4}g^{ab}\partial_{i}g_{ab}+\partial_{i}\phi\right)\left(g^{ji}\tilde{\partial}^{q}g_{pj}\right)\nonumber \\
 &  & +\frac{1}{2}\tilde{g}_{ip}\left(-\frac{1}{4}\tilde{g}_{ab}\tilde{\partial}^{l}\tilde{g}^{ab}+\tilde{\partial}^{l}\tilde{\phi}\right)\left(\tilde{g}_{kl}\partial_{j}\tilde{g}^{ik}\right)g^{jq}\nonumber \\
 &  & +\frac{1}{2}\tilde{g}_{ip}\left(-\frac{1}{4}g^{ab}\partial_{l}g_{ab}+\partial_{l}\phi\right)\left(g^{kl}\tilde{\partial}^{i}g_{jk}\right)g^{jq}\nonumber \\
 &  & +\frac{1}{2}\tilde{g}_{ip}\left(-\frac{1}{4}\tilde{g}_{ab}\tilde{\partial}^{l}\tilde{g}^{ab}+\tilde{\partial}^{l}\tilde{\phi}\right)\left(\tilde{g}^{ki}\partial_{k}g_{lj}\right)g^{jq}\nonumber \\
 &  & +\frac{1}{2}\tilde{g}_{ip}\left(-\frac{1}{4}g^{ab}\partial_{l}g_{ab}+\partial_{l}\phi\right)\left(g_{kj}\tilde{\partial}^{k}\tilde{g}^{li}\right)g^{jq}.
\end{eqnarray}

\noindent There also exist similar symmetries as in
$\star\mathcal{R}$

\noindent
\begin{eqnarray}
*\mathcal{R}_{\mathbf{1}\left(p\right)\mathbf{1}\left(q\right)} &
\underleftrightarrow{g^{ \bullet\bullet}\leftrightarrow
g_{\bullet\bullet},\quad\tilde{\partial}^{\bullet}
\leftrightarrow\partial_{\bullet},\quad\tilde{\phi}\leftrightarrow\phi}
&
*\mathcal{R}_{\mathbf{2}\left(p\right)\mathbf{2}\left(q\right)},\nonumber \\
*\mathcal{R}_{\mathbf{1}\left(p\right)\mathbf{2}\left(q\right)} &
\underleftrightarrow{g^{\bullet\bullet}\leftrightarrow
g_{\bullet\bullet},\quad\tilde{\partial}^{\bullet}\leftrightarrow
\partial_{\bullet},\quad\tilde{\phi}\leftrightarrow\phi}
& *\mathcal{R}_{\mathbf{2}\left(p\right)\mathbf{1}\left(q\right)}.
\end{eqnarray}

\noindent Similarly, from the metric ansatz (\ref{MetricAnsatz})
and calculation rules in section (4.1),

\noindent
\begin{eqnarray}
*\mathcal{R}_{\mathbf{2}\left(t\right)\mathbf{2}\left(t\right)} & = & \left(D-1\right)\frac{\dot{a}^{2}}{a^{2}}-\left(D-1\right)\frac{\ddot{a}}{a}+2\ddot{\phi}-\left(D-1\right)\frac{\dot{\tilde{a}}^{2}}{\tilde{a}^{2}}+\left(D-1\right)\frac{\ddot{\tilde{a}}}{\tilde{a}}-2\ddot{\tilde{\phi}},\nonumber \\
*\mathcal{R}_{\mathbf{2}\left(i\right)\mathbf{2}\left(i\right)} & = & \frac{1}{\tilde{a}^{4}}\left(-\left(D-1\right)\dot{\tilde{a}}^{2}+2\dot{\tilde{\phi}}\tilde{a}\dot{\tilde{a}}\right)-\left(-\left(D-1\right)\dot{a}^{2}+2\dot{\phi}a\dot{a}\right),\nonumber \\
*\mathcal{R}_{\mathbf{1}\left(t\right)\mathbf{2}\left(t\right)} &
= & 0.
\end{eqnarray}

\noindent Finally, combining two parts of the generalized Ricci
tensor,

\noindent
\begin{equation}
\mathcal{R}_{MN}=\star\mathcal{R}_{MN}+*\mathcal{R}_{MN}=0,
\end{equation}

\noindent  we get

\noindent
\begin{eqnarray}
\mathcal{R}_{\mathbf{2}\left(t\right)\mathbf{2}\left(t\right)} & = & -\left(D-1\right)\left(-\dot{H}+H^{2}\right)+2\ddot{\phi}+\left(D-1\right)\left(-\dot{\tilde{H}}+\tilde{H}^{2}\right)-2\ddot{\tilde{\phi}},\nonumber \\
\mathcal{R}_{\mathbf{2}\left(i\right)\mathbf{2}\left(i\right)} & = & \frac{1}{\tilde{a}^{2}}\left(\dot{\tilde{H}}-\left(D-1\right)\tilde{H}^{2}-2\dot{\tilde{\phi}}\tilde{H}\right)
-a^{2}\left(\dot{H}-\left(D-1\right)H^{2}-2\dot{\phi}H\right),\nonumber \\
\mathcal{R}_{\mathbf{1}\left(t\right)\mathbf{2}\left(t\right)} & =
& 0.
\end{eqnarray}

\noindent with symmetry

\noindent
\begin{eqnarray}
\mathcal{R}_{\mathbf{1}\left(p\right)\mathbf{1}\left(q\right)} & \underleftrightarrow{g^{\bullet\bullet}\leftrightarrow g_{\bullet\bullet},\quad\tilde{\partial}^{\bullet}\leftrightarrow\partial_{\bullet},\quad\tilde{\phi}\leftrightarrow\phi} & \mathcal{R}_{\mathbf{2}\left(p\right)\mathbf{2}\left(q\right)},\nonumber \\
\mathcal{R}_{\mathbf{1}\left(p\right)\mathbf{2}\left(q\right)} &
\underleftrightarrow{g^{\bullet\bullet}\leftrightarrow
g_{\bullet\bullet},\quad\tilde{\partial}^{\bullet}\leftrightarrow\partial_{\bullet},\quad\tilde{\phi}\leftrightarrow\phi}
& \mathcal{R}_{\mathbf{2}\left(p\right)\mathbf{1}\left(q\right)}.
\end{eqnarray}

\end{document}